\documentclass[preprint,aps]{revtex4}
\pdfoutput=1
\usepackage{dcolumn}% Align table columns on decimal point
\usepackage{bm}% bold math
\usepackage{graphicx}% Include figure files
\usepackage{epsfig}
\usepackage{amsmath}
\usepackage{amssymb}
\usepackage{mathrsfs}
\usepackage{ulem}
\usepackage{color}
\usepackage{hyperref}
\usepackage{float}
\usepackage{subfig}
\usepackage{tabularx}
\usepackage{multirow}
\allowdisplaybreaks

\begin{document}

\begin{flushright}
\vspace{-0.2cm}EFI-19-7\\
\end{flushright}

\title{Bounding the Charm Yukawa}

\vspace*{0.2cm}

\author{Nina M. Coyle$^a$, Carlos E.M. Wagner$^{a,b,c}$, and Viska Wei$^{a}$ 
\vspace{0.2cm}
\mbox{}
}
\affiliation{
\vspace*{0.5cm}
$^a$~\mbox{Physics Department and Enrico Fermi Institute, University of Chicago, Chicago, IL 60637}\\
$^b$  \mbox{Kavli Institute for Cosmological Physics, University of Chicago, Chicago, IL 60637}\\
$^c$ \mbox{High Energy Physics Division, Argonne National Laboratory, Argonne, IL 60439}\\
}

\begin{abstract}
The study of the properties of the observed Higgs boson is one of the main research activities in High Energy Physics.  Although the couplings of the Higgs to the weak gauge bosons and third generation quark and leptons have been studied in detail, little is known about the Higgs couplings to first and second generation fermions. In this article, we study the  charm quark Higgs coupling in the so-called $\kappa$ framework.  We emphasize the existence of specific correlations between the Higgs couplings that can render the measured LHC Higgs production rates close to the SM values in the presence of large deviations of the charm coupling from its SM value, $\kappa_c = 1$.  Based on this knowledge,  we update the indirect bounds on $\kappa_c$ through a fit to the  precision Higgs measurements at the LHC.  We also examine the limits on $\kappa_c$ arising from  the radiative decay $H \to J/\psi + \gamma$, the charm quark-associated Higgs production,  charm quark decays of the Higgs field, charge asymmetry in $W^{\pm} + H$ production, and differential production cross section distributions. Estimates for the future LHC sensitivity on $\kappa_c$ at the high luminosity run are provided. 
\end{abstract}

\maketitle

\section{Introduction}
The Standard Model (SM) of particle physics provides a renormalizable and  gauge invariant description of particle interactions.  It
therefore makes testable predictions which are being probed at high energy physics experiments~\cite{Tanabashi:2018oca}. No clear evidence of a departure
of the SM predicted behavior has been observed.  However, while the predicted gauge interactions have been tested with great
precision \cite{Group:2012gb,ALEPH:2005ab,LEP:2003aa,Schael:2013ita}, the tests of the interactions of the recently discovered Higgs boson have not yet reached the same level of accuracy.

The Higgs production at the LHC has been probed in many different channels and the rates are in agreement with the SM predicted
ones at a level of a few tens of percent \cite{ATLAS:2018doi,Aaboud:2018zhk,Sirunyan:2018koj}. Since in the SM those rates are mostly governed by the coupling of the Higgs to weak gauge 
bosons and third generation quarks, this  suggests that the observed Higgs production rates are governed by SM interactions
and that those couplings are within tens of percents of their SM predicted values. Global fits to the Higgs precision measurements
confirm this picture, showing no clear evidence of new physics coupled to the Higgs \cite{ATLAS:2018doi},\cite{Sirunyan:2018koj}. 

In spite of these facts, it is still very relevant to continue studying the properties of the Higgs boson in great detail.  First of all,
there could be deviations from the SM predictions at a level not yet probed by the LHC, which may reveal the presence of new physics at the weak scale. Second, the
couplings to the first and second generation of quarks and leptons have not been tested and deviations from their SM predicted
values may point towards a more complex mechanism of mass generation than the one present in the SM.  Third, there may be decays
of the Higgs bosons into exotic particles not yet detected by the LHC.  Last but not least, there may be
hidden correlations between the Higgs couplings  that may lead  to rates in agreement with the SM predicted ones, in spite of  
deviations of the couplings from the SM values. In this work, we shall present examples of such possible correlations. 

In this work, we shall study possible effects of the deviations of the charm-quark  Higgs coupling with respect to the SM value in the $\kappa$ 
framework \cite{LHCHiggsCrossSectionWorkingGroup:2012nn,Heinemeyer:2013tqa}, in which $\kappa_i$ characterize the ratio of a given coupling with respect to its SM value.  Large deviations of $\kappa_c$ from one affect the Higgs width 
and therefore its decay branching ratios, and therefore the couplings of the Higgs to gauge bosons and third generation fermions must be modified as
well in order to preserve the agreement with experimental observations. We shall study these modifications in detail and discuss their impact on the
determination of the charm quark  coupling to the Higgs boson. 

Let us emphasize that the $\kappa$ framework can not replace a more complete study of the Higgs properties based on higher order operators
coming from integrating out the new physics at the TeV~scale~\cite{Gupta:2014rxa,Contino:2013kra,Falkowski:2015fla,deFlorian:2016spz}. In particular, important effects related to 
for instance the energy dependence of the
form factors associated with these operators, or the correlation of the modification of the Higgs couplings with electroweak precision measurements,
are missed in the $\kappa$ framework.  However,  this framework is appropriate to obtain
an estimate of the possible sensitivity to unknown couplings, like
the one of the charm quark to the Higgs, where the current bounds are far from the SM values.  Moreover, the $\kappa$ framework is used by the
ATLAS and CMS collaborations and hence allows a direct comparison with the experimental results for values of $\kappa_c \simeq 1$.  

The article is organized as follows.  In section~\ref{sec:rates},  we shall determine
the spectific correlations between the Higgs couplings  that are necessary to keep the LHC Higgs production rates 
close to the SM ones. Using these results,  in Section~\ref{sec:precmeas} we shall study  the constraints that current precision Higgs measurement
impose on the Higgs couplings. In Section~\ref{sec:radiative} we shall discuss the bounds on the Higgs couplings coming from the measurement of
radiative decays of the Higgs boson into charmonium states.  Finally, in Section~\ref{sec:LHCfut} we shall discuss the impact of LHC Higgs production 
and decay rates 
induced by the charm coupling. We reserve Section~\ref{sec:Conc} for our conclusions.

\section{Best-fit values on Higgs rates}
\label{sec:rates}

The rate of a Higgs production and decay process relative to the Standard Model rate is represented by the signal strength $\mu_{if}$, where
\begin{equation}
	\mu_{if} = \frac{\sigma_i \times B_f}{(\sigma_i \times B_f)^{SM}},
\end{equation}
is the ratio of the  product of the Higgs production cross section $\sigma_i$ in a given $i$-channel and its decay branching ratio $B_f$  in a given $f$-channel  to their SM 
predicted values. 
Within the $\kappa$ framework, the quantity $\sigma_i \times B_f$ can be obtained by a simple rescaling of each couplings by a corresponding factor $\kappa$  and it is therefore expressed as
\begin{equation}
	\sigma_i \times B_f = \kappa_{r,i}^2 \sigma^{SM}_{i} \times \frac{\kappa_{f}^2 \Gamma^{SM}_{f}}{\Gamma_H}
\end{equation}
where $\kappa_{r,i}$ is associated with the relevant Higgs coupling governing the $i$ production mode, while $\kappa_f$ is associated with the Higgs coupling
governing the decay into particles $f$, with SM partial width $\Gamma^{SM}_f$.  The total Higgs width $\Gamma_H$ is hence calculated as
\begin{align} 
	\Gamma_H =& \Gamma^{SM}_{H} \big( \kappa^2_b B^{SM}_{bb} + \kappa_W^2 B^{SM}_{WW} + \kappa_g^2 B^{SM}_{gg} + \kappa_{\tau}^2 B^{SM}_{\tau\tau} + \kappa_{Z}^2 B^{SM}_{ZZ} + \kappa_{c}^2 B^{SM}_{cc} + \kappa_{\gamma}^2 B^{SM}_{\gamma\gamma} \nonumber \\
	&+ \kappa^2_{Z\gamma} B^{SM}_{Z\gamma} + \kappa^2_s B^{SM}_{ss} + \kappa^2_{\mu} B^{SM}_{\mu\mu} \big) /(1-B_{BSM}) \\
	\equiv& \;  \Gamma^{SM}_H \kappa^{2}_{H},
\end{align}
where  $B^{SM}_f$ is the decay branching ratio in a given $f$ channel within the SM and $B_{BSM}$ is the branching ratio of the Higgs decay into beyond the SM particles. 
Here and in the following we have treated the loop-induced coupling of the Higgs to gluons and photons as independent quantities, and therefore not restricted to the loop contributions of only SM particles. 

The rates relative to the SM ones in this framework are therefore written as
\begin{equation} \label{eq:mukappas}
	\mu_{if} = \frac{\kappa^2_{r,i} \kappa^2_{f}}{\kappa^2_H}.
\end{equation}
It is important to remark that, considering the photon and gluon couplings as independent variables, the Higgs production rates in the standard channels (gluon fusion,
weak boson fusion and associated production of the Higgs with gauge bosons, top and bottom pairs)
are not affected in any relevant way by the charm Yukawa coupling.  However, the decay
rates are affected in a clear way by a modification of $\kappa_c$. Indeed,
the value of $\kappa_c$ influences $\kappa^2_H$, therefore decreasing the rates of the observed processes by increasing the total width. Because we are interested in finding an upper bound on $|\kappa_c|$, we will not include a non-zero $B_{BSM}$ term, which would have the same effect on the rates as increases in $|\kappa_c|$.

In order to obtain bounds on $|\kappa_c|$, we examine how well the measured rates can be fitted for increasing values of the charm Yukawa. The fit includes the most recent 13 TeV results for the observed rates from ATLAS, contained in Refs. \cite{ATLAS:2018doi} and \cite{Aaboud:2018zhk}, and CMS, contained in Ref.~\cite{Sirunyan:2018koj}. We fit to a weighted average of the experiments' measurements. The free parameters included in our fit are \{$\kappa_b$, $\kappa_W$, $\kappa_t$, $\kappa_Z$, $\kappa_{\tau}$, $\kappa_g$, $\kappa_{\gamma}$\} with $\kappa_c$ as an input.  %We shall ignore the small contributions to the width induced by decays into strange quarks and muons as well as all fist generation particles, as well as the decays into $Z\gamma$ final states, which can only contribute for extremely large values of their respective $\kappa$'s. 
%As mentioned above, we treat $\kappa_g$ and $\kappa_{\gamma}$ as free independent parameters, allowing for unknown contributions to these effective couplings from BSM loops. 
We examine three scenarios: one in which the values of $\kappa_W$ and $\kappa_Z$ are unconstrained, one based on estimates of the bounds coming from precision electroweak measurements,
and the last in which $\kappa_W, \kappa_Z \leq 1$. The latter situation is less general but is well motivated by theory. We take $\kappa_{Z\gamma}$, $\kappa_{s}$, and $\kappa_{\mu}$ to be equal to 1 since they are not directly involved in the fitted processes and may contribute in a relevant way to the total width only for extreme values of their respective $\kappa$ values. 

While performing a fit to the Higgs couplings based on only the currently measured production rates, we found that no meaningful bound on $\kappa_c$ could be obtained. 
The reason for this behavior is the existence of a flat direction in the fit 
for which all $\kappa$'s increase along with the increasing $\kappa_c$.   This fact was already  emphasized for instance by the authors of 
Refs.~\cite{Zeppenfeld:2000td,Djouadi:2000gu,Duhrssen:2004cv,Belanger:2013xza}, 
who noticed that no additional, unobserved decays may be constrained by a simple fit to the observed production and decay rates. Although this observation
was related to a possible invisible decay width, it can also be applied to the case of unobserved decays into charm quarks, in which case, by a suitable modification of the
$\kappa_i$, 
the observed rates can be modeled equally well for any value of $\kappa_c$.  To see this, we can write down the rate for a given observed process as
\begin{align} 
\label{eq:muifk}
	\mu_{if} = %\frac{ \sigma_i^{SM} \kappa^2 \times \Gamma^{SM}_{f} \kappa^2}{\Gamma_{H}} \nonumber 
	\frac{  \kappa^4 } { \kappa^2(1 - B_{cc}^{SM}) + \kappa_c^2 B_{cc}^{SM} }
\end{align}
where since all $\mu_{if} \simeq 1$ 
we have considered that all non-charm Higgs couplings scale together by a single $\kappa$ value. If we require the signal strengths $\mu_{if} $ to
be given by a value $\mu$, 
Eq.~(\ref{eq:muifk}) provides a quadratic equation on $\kappa^2$. The solution to this quadratic equation leads to a correlation between the  necessary values of the generic $\kappa$
and $\kappa_c$, namely
\begin{align}
%	\frac{ \kappa^4}{ \kappa^2 (1-B_{cc}^{SM}) + \kappa_c^2 B_{cc}^{SM} } = \mu \\
	\kappa^2 = \frac{(1-B_{cc}^{SM})\mu}{2} + \frac{ \sqrt{(1-B_{cc}^{SM})^2 \mu^2+4\mu B_{cc}^{SM}\kappa_c^2} }{2} .
\end{align}
Since, as stressed before, the observed rates are all within tens of percents of the SM values, one should require $\mu \approx 1$ in order to obtain agreement with the precision Higgs
measurements. Therefore, given that $B_{cc}^{SM} \simeq 0.03$, an unconstrained fit to all couplings will lead to the following approximate correlation between the Higgs couplings
\begin{equation}
\label{eq:corr}
	\kappa^2 \approx \frac{0.97}{2} + \frac{ \sqrt{(0.97)^2+0.12\kappa_c^2}}{2}
\end{equation}
which clearly has a solution for all real $\kappa_c$.

\section{Constraints on $\kappa_c$ from Higgs precision measurements}
\label{sec:precmeas}

The existence of the flat direction described in Eq.~(\ref{eq:corr}) implies that no contraints on the $\kappa_c$ values may be obtained by 
considering only the current Higgs precision measurements. Additional constraints are therefore necessary to put a bound on $\kappa_c$.  In this section, we
shall describe the constraints imposed by the bounds on the total Higgs width, the ones coming from precision electroweak measurements,
and finally the ones coming from the theoretical prejudice that, in most extensions of the SM, $\kappa_V \leq 1$. 

In all cases we perform a fit to $\kappa_c$ marginalizing over all the other couplings. The channels included in the fit are shown in Table \ref{tab:channels}. In addition to the individual decay channels listed in the table, we also include the combined results for each given production mode. We combine the ATLAS and CMS results given in \cite{ATLAS:2018doi,Aaboud:2018zhk,Sirunyan:2018koj} by a weighted average, weighting by the squared inverse of the respective 1$\sigma$ uncertainties. The uncertainty in the combined observation is given by
\begin{equation}
	\sigma^{comb.}_{if} = \frac{1}{\sqrt{ 1/(\sigma_{if}^{ATLAS})^{2} + 1/(\sigma_{if}^{CMS})^{2}}}
\end{equation}
where $\sigma_{if}$ indicates the uncertainty in the corresponding observed value of $\mu_{if}$.

\begin{table}[H]
	\centering
	\begin{tabular}{| c | l || c | l |}
		\hline 
		Production mode & Decay mode & Production mode & Decay mode \\ \hline
		\multirow{4}{*}{ggF} & $H\to\gamma\gamma$ & \multirow{4}{*}{VH} & $H \to \gamma\gamma$ \\
		& $H \to ZZ$ & & $H \to ZZ$ \\
		& $H \to WW$ & & $H \to bb$ \\
		& $H \to \tau \tau$ & & \\ \hline
		\multirow{4}{*}{VBF} & $H \to \gamma\gamma$ & \multirow{4}{*}{ttH} & $H \to \gamma\gamma$ \\\
		& $H \to ZZ$ & & $H \to VV$  \\
		& $H \to WW$ & & $H \to \tau\tau$  \\
		& $H \to \tau\tau$ & & $H \to bb$ \\ \hline			 			
	\end{tabular}
	\caption{The production and decay channels included in the fit over $\kappa$'s. We also include the combined results for each production mode.}
	\label{tab:channels}
\end{table}

The $\chi^2$ value for a given fit is calculated as
\begin{equation}
	\chi^2 = \sum_{if} \frac{ (\mu_{if}(\kappa) - \mu^{obs}_{if})^2}{\sigma_{if}^2}
\end{equation}
where $\mu_{if}(\kappa)$ represents the calculated value of $\mu_{if}$, using Eq.~(\ref{eq:mukappas}), for the given set of $\kappa$'s. We find the best fit at each $\kappa_c$ by minimizing the value of $\chi^2$ for the given $\kappa_c$.

In the cases where $\kappa_V$ is constrained, we obtain a 95\% CL bound by placing a limit on $\Delta \chi^2$ relative to the best fit at $\kappa_c=1$. In order to identify the appropriate $\Delta \chi^2$ cut, we performed a principle component analysis \cite{PCA_1,PCA_2} on a centralized data set of $\{\kappa_b, \kappa_W, \kappa_t, \kappa_\tau, \kappa_Z, \kappa_{\gamma}, \kappa_g\}$ for $\kappa_c \in [1.0,4.0]$, for $\kappa_V \leq 1$. We converted the 7-dimensional correlated $\kappa$ data into a set of uncorrelated principle components, and observed that the $99\%$-dominant principle component is an approximately equally-weighted linear combination of $\{\kappa_b, \kappa_t, \kappa_\tau, \kappa_{\gamma}, \kappa_g\}$. $\kappa_W$ and $\kappa_Z$ contribute trivially to the principle direction due to the constraint $\kappa_V \leq 1$. Thus we treat $\{\kappa_b, \kappa_W, \kappa_t, \kappa_\tau, \kappa_Z, \kappa_{\gamma}, \kappa_g\}$ as one fit parameter. Including the fit parameter coming from $\kappa_c$, our $\chi^2$ fit is effectively a 2-parameter fit. As a result, we will employ a $95\%$ CL cut corresponding to $\Delta\chi^2=5.99$. 

\begin{figure}[H]
	\centering
	\includegraphics[width=0.7\textwidth]{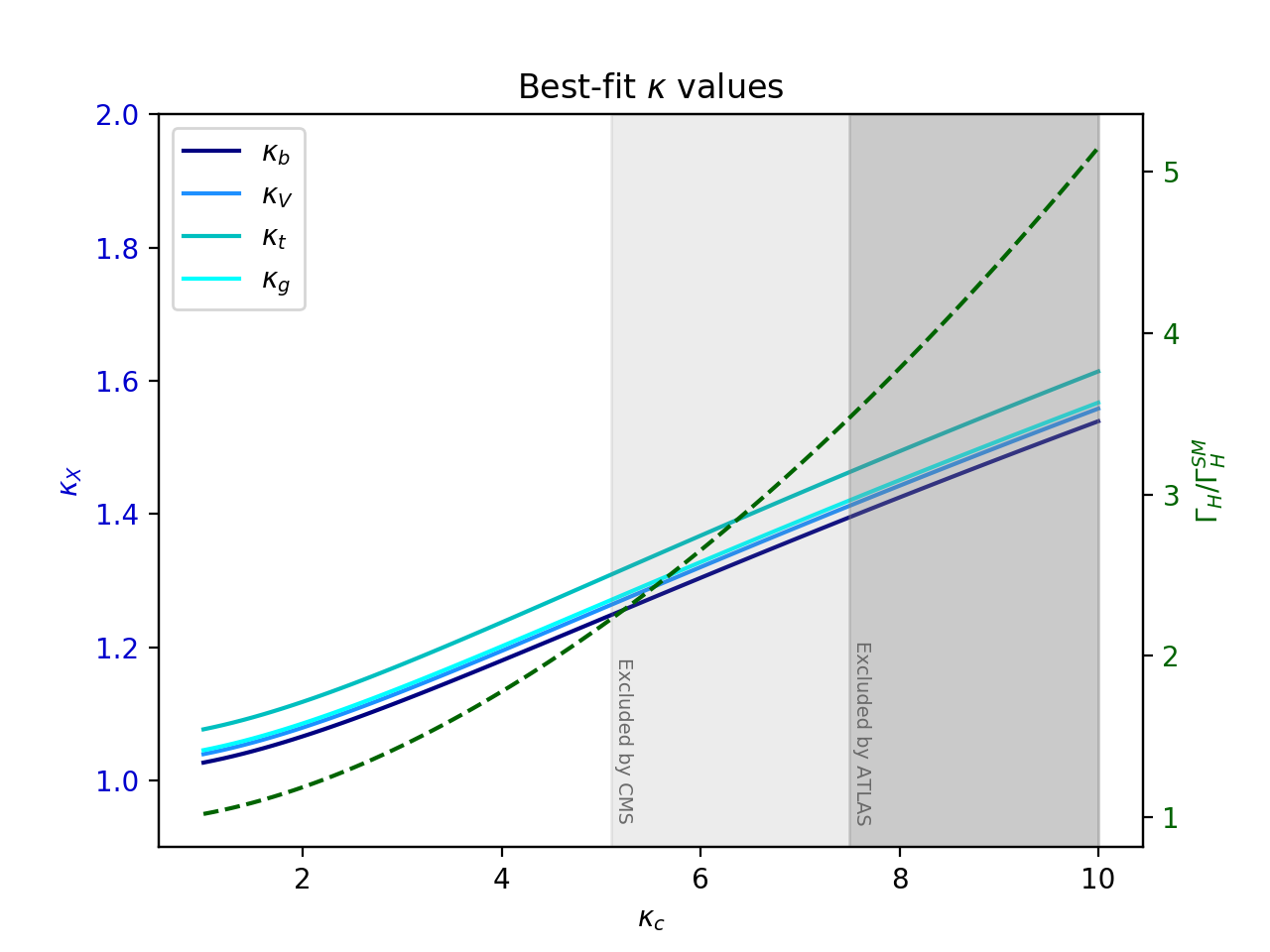}
	\caption{Plots of the best-fit values of $\kappa$'s, represented by solid lines, to the precision rate measurements $\mu_{if}$. The grey regions are excluded by constraints on the total Higgs width, which is normalized to the SM value and represented by a dashed line.}
	\label{flatdir}
\end{figure}

\subsection{Higgs decay width}

The  increase in all $\kappa$'s following the flat direction described in Eq.~(\ref{eq:corr}) leads to an increase in the total width $\Gamma_H$, and one may therefore place a bound on $|\kappa_c|$ using 
bounds on the Higgs width. ATLAS and CMS have performed maximum likelihood fits using on-shell and off-shell $H\to ZZ$ measurements to obtain a bound on the total Higgs width; they find 
\begin{align}
\Gamma_H &< 14.4~{\rm MeV} \;\;\;\;\; \text{(ATLAS)} \nonumber \\
\Gamma_H &< 9.16~{\rm MeV} \;\;\;\;\;\; \text{(CMS)}
\end{align} 
or $\Gamma_H / \Gamma^{SM}_H < 3.5$ and $\Gamma_H / \Gamma^{SM}_H < 2.2$, respectively, at 95\% CL \cite{Aaboud:2018puo, Sirunyan:2019twz}. It is necessary to note that these limits are obtained by making certain assumptions, in particular that the $\kappa$ values do not depend on the momentum transfer of the Higgs production mechanism and that $\kappa_{V} = \kappa_{g}$. Because $\kappa_V$ and $\kappa_g$ naturally have nearly equal values in the best fits, this second condition is indeed approximately satisfied. 

We perform a $\chi^2$ fit to the LHC measurements of  all measured signal strengths $\mu_{if}$, Eq.~(\ref{eq:mukappas}), for increasing values of $\kappa_c$ 
and find that the 95\% C.L. limits on the Higgs width lead to a bound of $|\kappa_c| < 7.5$ from ATLAS and $|\kappa_c| < 5.1$ from CMS.
Figure \ref{flatdir} shows a plot of the best-fit $\kappa$'s for increasing $\kappa_c$, and indicates the regions for which the total Higgs width, represented by the dashed-line, exceeds the current bounds. The spread in values for the various $\kappa$'s arises from the differences in individual rate measurements.

\subsection{Precision Electroweak Measurements}

It is also worth noting that the necessary increases in all $\kappa$ values to be consistent with the Higgs production rates result in $\kappa_V>1$. In particular, for $|\kappa_c|=7.5$ the least-squares fit gives values of $\kappa_W = 1.42$ and $\kappa_Z=1.38$, which are consistent with the approximate flat direction values given by Eq.~({\ref{eq:corr}).
 These large values for $\kappa_V$ result in divergences in electroweak precision parameters which are not canceled by the Higgs contribution, as they are in the SM. In this case one would require an extension of the SM which cancels the divergent contributions to the precision measurement variables.
One can replace the divergence by a parametric logarithmic dependence on an effective cutoff that
characterizes the new physics. In such a case, for instance, if one assumes a cutoff scale of the order of $\Lambda = 3$~TeV, a fit to the precision electroweak measurements leads to a value of $\kappa_V = 1. 08 \pm 0.07$ \cite{Falkowski:2013dza}.  Since $\kappa_V$ is now constrained to values lower than the ones necessary to reach the bounds on the Higgs width, there will be a stronger upper bound on $\kappa_c$. 

 In order to find a bound on $\kappa_c$ from this limit on $\kappa_V$, we include the deviation of $\kappa_V$ from $\kappa_V = 1.08$ in the calculation of $\chi^2$ and perform a $\chi^2$ fit for increasing $\kappa_c$. We examine the $\Delta \chi^2$ relative to the fit at $\kappa_c=1$. Performing a fit to the Higgs rates using 
this constraint on $\kappa_V$, one obtains $|\kappa_c| < 4.9$. Observe, however, that this bound depends on specific assumptions about the new physics scale.

\subsection{Constrained $\kappa_V$}

\begin{figure}[H]
	\centering
	\includegraphics[width=0.7\textwidth]{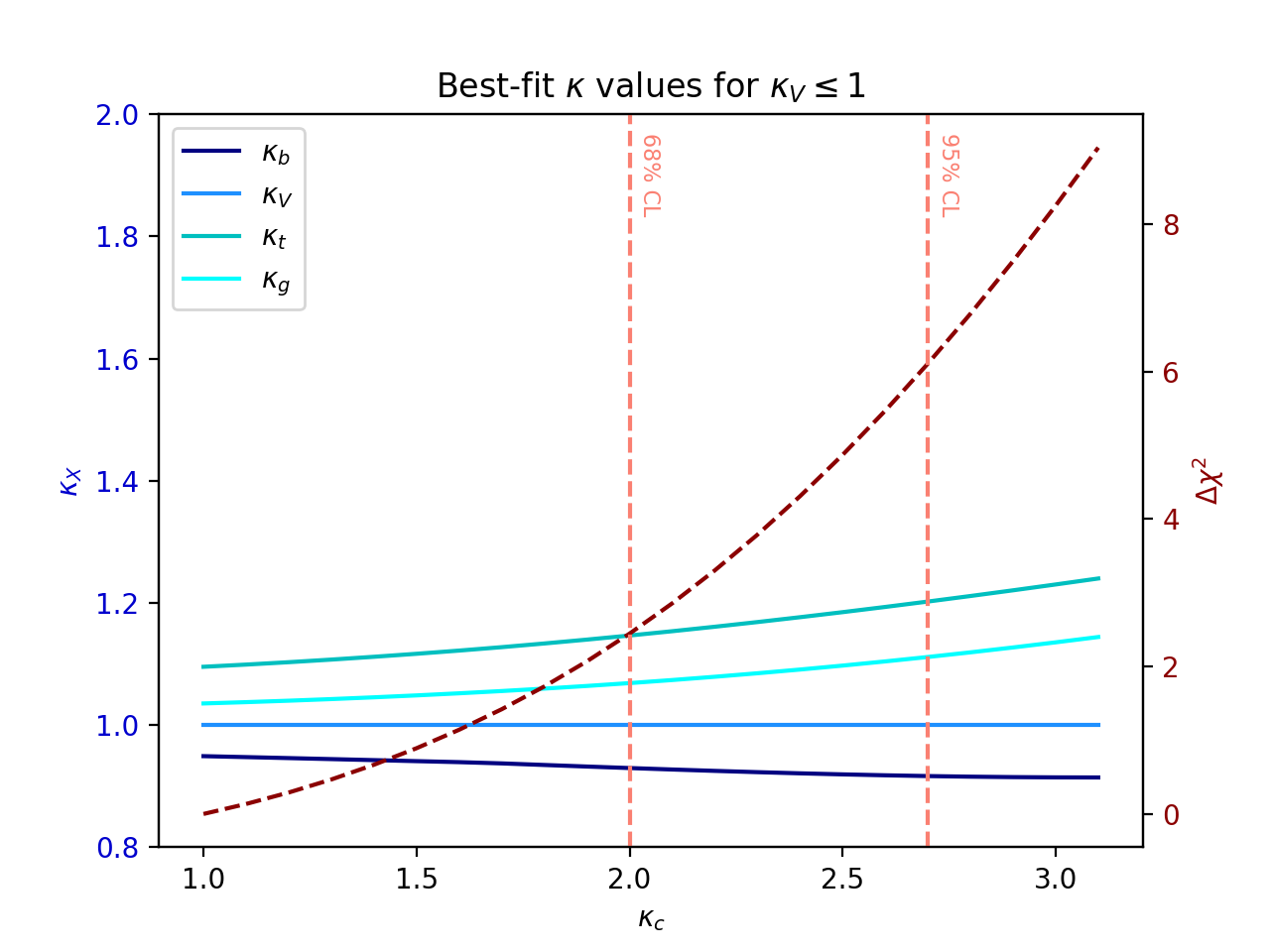}
	\caption{Plots of the best-fit values of $\kappa$'s for $\kappa_V \leq 1$. Although we plot $\kappa_W$ and $\kappa_Z$ together as $\kappa_V$, the two differ very slightly due to the differences in the W and Z rate measurements. The dashed line represents the $\Delta \chi^2$ of the fit at a given $\kappa_c$ relative to the $\chi^2$ of the fit at $\kappa_c=1$.}
	\label{constrained}
\end{figure}

In this third scenario, the flat direction is removed by constraining $\kappa_W, \kappa_Z \leq 1$. This constraint is well motivated, as models with extended Higgs sectors such as a 2HDM will typically include couplings to the weak gauge bosons lower than the SM values for the individual Higgs particles. Similarly to the previous case, the $\kappa$'s cannot increase uniformly to maintain the same relative strengths, so we expect that the fit will become less accurate as the total width increases through $\kappa_c$. As in the previous section, we obtain a 95\% CL bound on $\kappa_c$ by identifying the value of $\kappa_c$ for which the least-squares fit has $\Delta \chi^2= 5.99$ relative to the best fit at $\kappa_c=1$. We find a bound of $|\kappa_c| < 2.7$ at 95\% CL. Figure \ref{constrained} shows a plot of the behavior of the best-fit $\kappa$'s, represented by solid lines,  for increasing $\kappa_c$ along with the value of $\Delta \chi^2$, represented by a dashed line.

\subsection{Future prospects for the HL-LHC}
We can examine these cases for the HL-LHC, for which the projected uncertainties of the rate measurements have been examined for ATLAS~\cite{ATLAS_mu} and CMS~\cite{CMS:2018qgz}. We update the 1$\sigma$ uncertainties used in our $\chi^2$ fit using the combined expected errors quoted in the two studies. In the case of the width constraint, if only the on-shell rate measurements are considered, the bound on $|\kappa_c|$ remains approximately the same, as the $\kappa$ values along the flat direction are similar regardless of the uncertainties in $\mu_{if}$. However, the width bound is also expected to improve with higher luminosity. According to an ATLAS study of off-shell Higgs to ZZ measurements for the HL-LHC \cite{ATLAS_width}, assuming the observed on-shell and off-shell rates are equal to the SM prediction, the expected determination of $\Gamma_H$ with 3 ab$^{-1}$ is
\begin{equation}
	\Gamma_H = 4.2^{+1.5}_{-2.1} \text{ MeV}
\end{equation}
or $\Gamma_H/\Gamma_H^{SM} = 1.0^{+0.4}_{-0.5}$. Requiring that the width remains consistent with this expectation corresponds to a bound of $|\kappa_c| \lesssim 3.0$. 

The projected constraints for $\kappa_V \leq 1$ depend somewhat on the values of $\mu_{if}$ one uses in the fit. The projection studies use $\mu_{if}=1$ for all initial and final states to estimate the percent uncertainty on each measurement.  An alternative method is to adjust the percent uncertainty to the expected HL-LHC values but use the current measurements; this method is not ideal, as limiting the uncertainties without changing the values of $\mu_{if}$ is unlikely to accurately reflect the HL-LHC results. However, the comparison of the bounds on $\kappa_c$ obtained in the two scenarios provide a good picture of the likely constraints on this quantity.
For $\mu_{if}$ equal to the current measurements, we find an expected bound of $|\kappa_c| < 2.2$. On the other hand, for $\mu_{if} = 1$, the expected bound is given by $|\kappa_c| < 2.1$. We therefore expect the HL-LHC to provide an indirect limit of $|\kappa_c| \lesssim 2.1$ in the $\kappa_V \leq 1$ case.

\section{Radiative Higgs Decay to $J/ \psi$}
\label{sec:radiative}

Radiative decays of the Higgs boson into charmonium states are known to provide a sensitive probe of the charm coupling, and have been previously examined in this context in \cite{Perez:2015aoa, Perez:2015lra, Koenig:2015pha, Bodwin:2013gca}.
This is due to the fact that the charm-coupling induced rates interfere with those induced by the top and W couplings in a well-defined
way.  For instance, the width for $H \to J/\psi + \gamma$ is given by \cite{Bodwin:2014bpa}
\begin{equation} 
\label{eq:raddec}
	\Gamma(H \to J/\psi + \gamma) = |(11.9 \pm 0.2) \kappa_{\gamma} - (1.04 \pm 0.14)\kappa_c|^2 \times 10^{-10} \text{ GeV} 
\end{equation}
where the first term arises from the amplitude which contains no dependence on $\kappa_c$ and the second from the  $\kappa_c$-dependent amplitude. Plugging in $\kappa_{\gamma},\kappa_c=1$ and $\Gamma_H^{SM} = 4.195 \times10^{-3}$~GeV gives the SM value for the branching ratio as 
\begin{equation}
\label{eq:SMrad}
BR^{SM}(H \to J/\psi + \gamma) = 2.79 \times 10^{-6}. 
\end{equation}
The current bound on this process is 
\begin{equation}
\sigma \times BR(H \to J/ \psi + \gamma) < 19~{\rm fb} . 
\end{equation}
at 95\% CL. Assuming the SM production cross section~\cite{Aaboud:2018txb}, this limit corresponds to 
\begin{equation}
\label{eq:boundrad}
BR(H \to J/\psi + \gamma) < 3.5 \times 10^{-4}.
\end{equation}

Since the production cross section depends on the values of $\kappa's$, which should increase together with $|\kappa_c|$ in order to keep agreement with the Higgs production rates,
this bound on the branching ratio is only useful for moderate values of $\kappa_c$, for which $\sigma_{H} \approx \sigma_{H}^{SM}$. However, the bound on the branching ratio is two orders of magnitude larger than the SM branching ratio, and therefore cannot currently probe moderate values of $\kappa_c$. Additionally, the branching ratio displays asymptotic behavior for large $\kappa_c$, as there are also $\kappa_c$-dependent enhancements of the Higgs total width. For large $\kappa_c$, the approximate expression for the branching ratio along the flat direction is given by
\begin{equation}
\label{eq:asympt}
	BR(H \to J/\psi + \gamma) \approx \frac{(5 |\kappa_c|^{1/2}-1.04 \kappa_c)^{2} \times 10^{-10}\ {\rm GeV}}{ (0.16 |\kappa_c| + 0.03 \kappa_c^2)\times \Gamma_H^{SM}}.
\end{equation}

Figure \ref{BR_JPsi} shows a plot of the behavior of this Higgs radiative decay branching ratio along the flat direction as well as with SM-like values for the other couplings. %The two cases have asymptotic values of $8.6 \times 10^{-7}$ and $8.0\times 10^{-7}$, respectively. 
We stress again that setting the other Higgs couplings to SM values for large $|\kappa_c|$ does not align well with rate measurements from the LHC, and it is therefore more instructive to examine the flat direction for large $|\kappa_c|$. In both cases, the branching ratio peaks at moderate negative values of $\kappa_c$,  at a maximum value of approximately $4\times10^{-6}$, two orders of magnitude below the current limit for SM production rates.

\begin{figure}[H]
	\centering
	\includegraphics[width=0.7\textwidth]{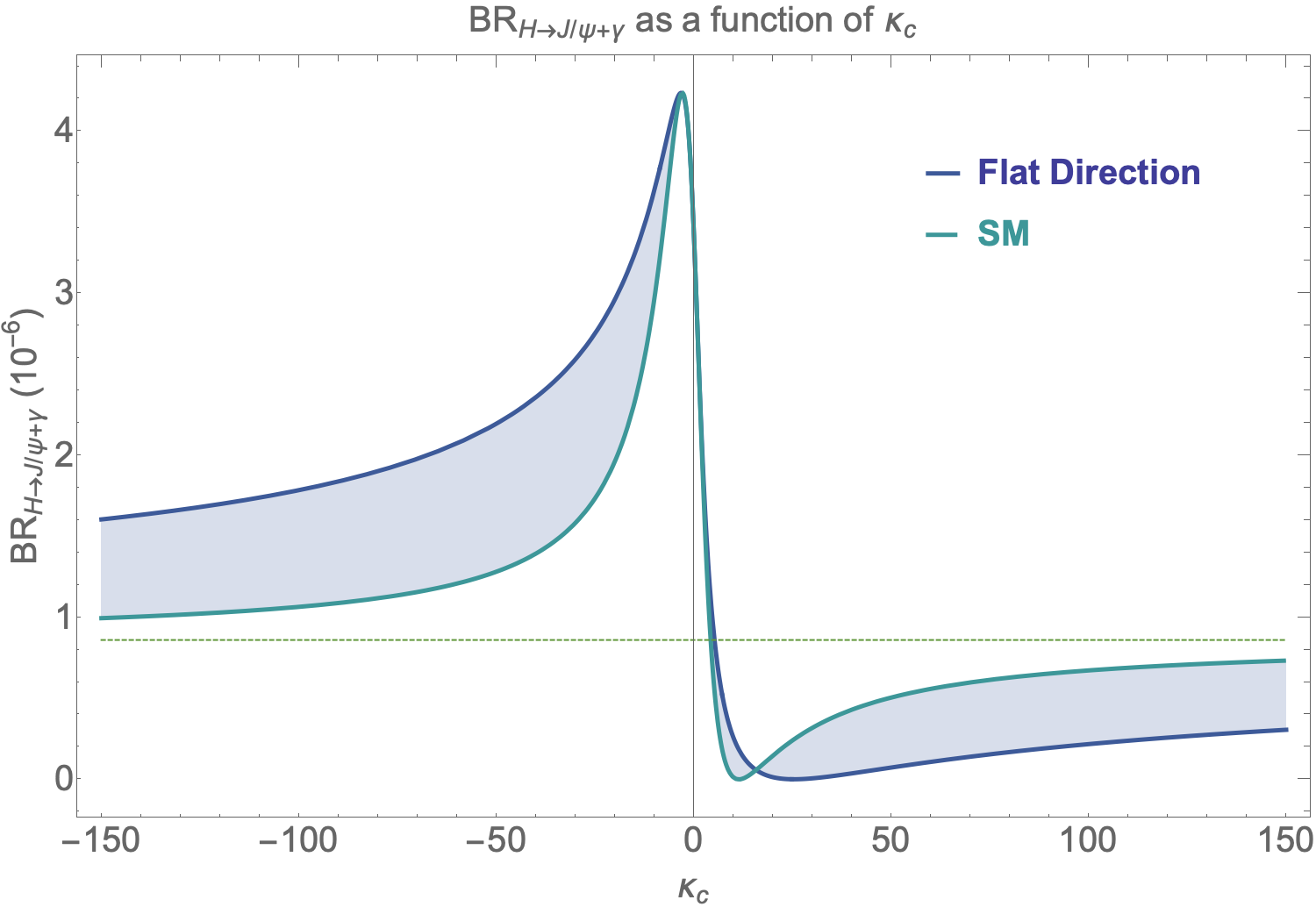}
		\caption{Plot of the branching ratio of $H \to J/\psi + \gamma$ varying along the flat direction (dark blue) and with other Higgs couplings fixed to SM values (light blue). The expected asymptote of approximately $8 \times 10^{-7}$ is indicated by the green dashed line. }
	\label{BR_JPsi}
\end{figure}

%As noted above, the quoted limit on the branching ratio is obtained by assuming SM Higgs production. However, since for large values
%of $\kappa_c$ all $\kappa$'s should increase in order to keep agreement with the observed Higgs rate, the production cross section
%will also increase and this effect must be taken into account while considering the $\kappa_c$ constraints. Moreover,  
%since the charm quark is a relevant proton constituent at low values of $x$, the Higgs production may be induced in proton
%collisions via its coupling to the charm quark.  In fact, 
%the Higgs production cross section will increase significantly for large $\kappa_c$, and this enhancement must be also taken into account. 

Given the non-SM production rate and asymptotic behavior of the branching ratio for large $\kappa_c$, we consider the limit on $\sigma \times BR$ rather than only the branching ratio. The production cross section increases due to both $\kappa^2$ enhancements given by Eq.~(\ref{eq:corr}) as well as $\kappa_c$-dependent processes such as $c\bar{c}H$ production, which become relevant for very large $\kappa_c$. We fit data produced with MadGraph 5 \cite{Alwall:2014hca} at leading order to obtain an expression for the approximate scaling of $\sigma_{c\bar{c}H}$ for large $\kappa_c$ at 13 TeV, which is given by
\begin{equation} \label{ccH}
	\sigma_{c\bar{c}H} \approx \left| 5.24\times10^{-2} + 2.76\times10^{-2} \kappa_c - 5.45\times10^{-6} \kappa_c^2 + 1.30\times10^{-6} \kappa_c^3  \right|^{2} {\rm pb}
\end{equation}
We also include contributions to $VH$ production from $c + \bar{c}/\bar{s}$ initial states. Figure \ref{sig_JPsi} shows a plot of $\sigma \times BR_{J/\psi}$ in fb for the flat direction.

Considering properly the rate, instead of just the radiative decay branching ratio, a limit can now be set for very large values of $\kappa_c$. By the end of the HL-LHC, the expected 95\% CL upper bound on $\sigma \times BR(H \to J/\psi + \gamma)$ from ATLAS is approximately 3 fb \cite{Aaboud:}. We therefore expect this process to place a limit of $\kappa_c \in [-180,330]$ at the HL-LHC for the flat direction. This limit is two orders of magnitude larger than those from other HL-LHC prospects discussed previously. %although it does provide a limit that is less model-dependent than the ones discussed in the previous section.  
A strong improvement, of  an order of magnitude of the present expected sensitivity, would be necessary for this channel to provide a competitive bound on $\kappa_c$. 

\begin{figure}[H]
	\centering
	\includegraphics[width=0.7\textwidth]{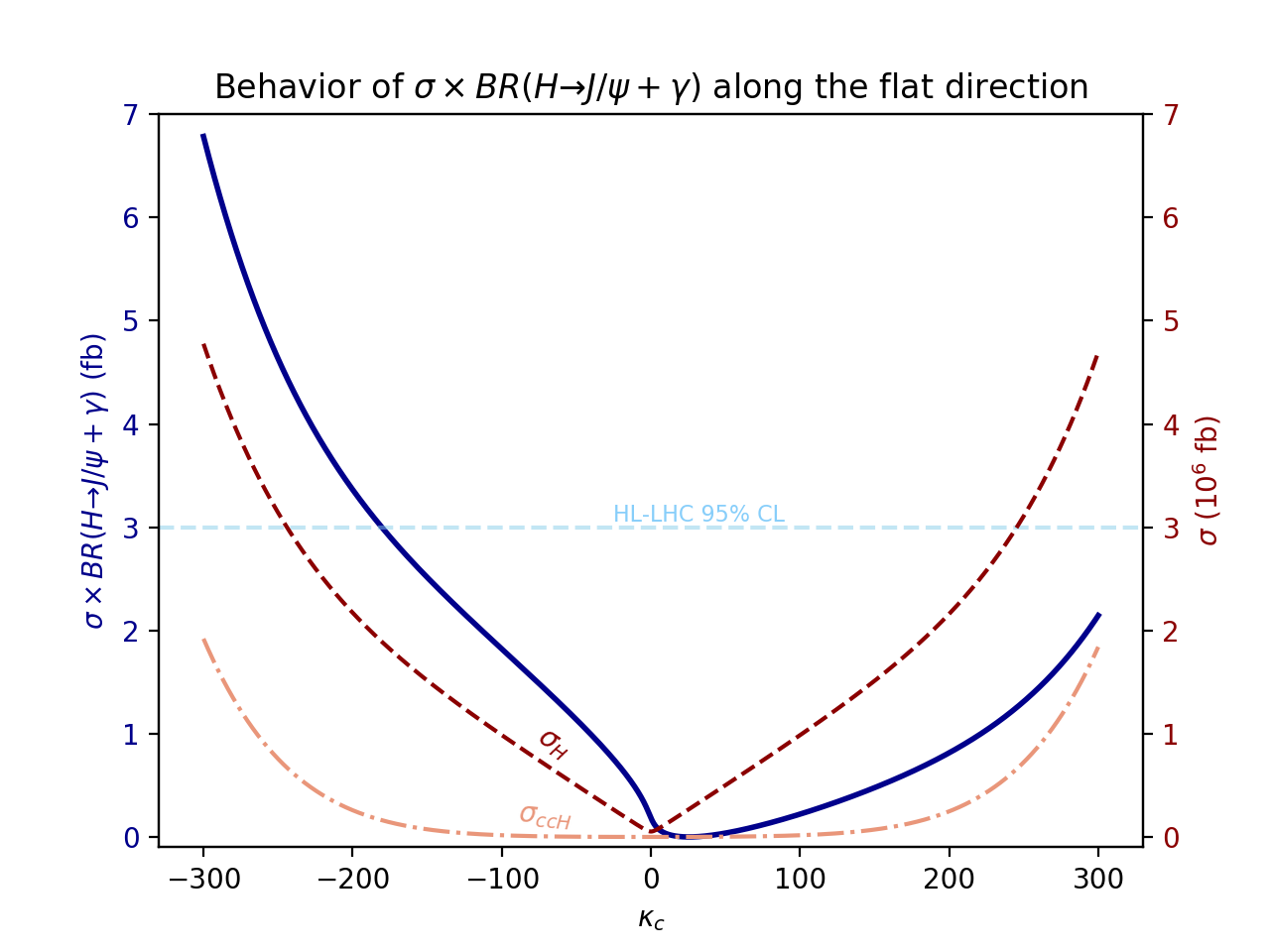}	
	\caption{Plot of $\sigma \times BR(H \to J/ \psi + \gamma)$ for the flat direction. The blue line indicates $\sigma \times BR$ in fb, while the pink dot-dashed (dashed) lines indicate the $c\bar{c}H$ (total) production cross section in fb. The dashed grey line shows the expected HL-LHC 95\% CL bounds.}
	\label{sig_JPsi}
\end{figure}

The authors of Ref. \cite{Bodwin:2014bpa} have updated the partial width expression with a new approach to the resummation of logarithms, and quote a new width of \cite{Bodwin:2016edd}
\begin{equation}
	\Gamma(H \to J/\psi + \gamma) = \left| (11.71 \pm 0.16)\kappa_V - ( (0.627^{+0.092}_{-0.094}) + i(0.118^{+0.054}_{-0.054}))\kappa_c \right|^2.
\end{equation}
This expression has a reduced dependence on $\kappa_c$, and therefore gives even weaker bounds on $\kappa_c$ than those found above.

It is important to note that such large values of $\kappa_c$ encounter strong experimental and theoretical issues. On the one hand, following the flat direction in order to retain consistency with precision Higgs measurements leads to large values of the top-quark coupling to the Higgs $g_{ht\bar{t}}$. In particular, for values of $\kappa_c \gtrsim 100$ one requires values of $\kappa_t \gtrsim 17$. In this case, the value of $g_{ht\bar{t}}^2$ is greater than $4\pi$, and a perturbative examination of the Higgs sector becomes unreliable. One may attempt to avoid this issue by fixing $\kappa_t$ to be less than a certain value, in which case the Higgs rates would become inconsistent with those observed at the LHC. We therefore note that such large values of $\kappa_c$ are problematic for either LHC Higgs rates or perturbativity concerns. Moreover, as stressed in Section \ref{sec:precmeas}, unless a very particular momentum dependence of the effective couplings is present, large values of 
$\kappa_c \gg 10$ would lead to a value of the Higgs width that is under strong tension with current LHC measurements. 

%The HL-LHC limit from $ZH \to l^{+} l^{-} c \bar{c}$, discussed in the next section, provides a much stronger limit from another model-independent search method.

%As stressed previously, $c\bar{c}H$ production dominates the total Higgs production rate for such large values of $\kappa_c$, which will affect the Higgs phenomenology. It is important to note that the analytical flat direction discussed previously will then be broken, as $\kappa_c$ now contributes to production rates. The effects of such large $\sigma_{ccH}$ would appear in the observed rate of other production processes before the $J/ \psi$ bounds are met. More precisely, the flat direction derived above holds for $\sigma_H \gg \sigma_{ccH}$; this is satisfied up to approximately $\kappa_c \approx 80$. Therefore, the $H \to J / \psi + \gamma$ channel does not provide a promising independent bound on $\kappa_c$, as phenomenological effects of large $\kappa_c$ will appear at values of $\kappa_c$ lower than the expected HL-LHC bounds. 

\section{Higgs Production Rates induced by the charm Higgs coupling}
\label{sec:LHCfut}

%The radiative decays of the Higgs are not the only production or decay processes sensitive to the charm quark Yukawa coupling. 
As stressed before, Higgs production  may be induced in proton
collisions via its coupling to the charm quark.  Moreover, the Higgs boson may decay into charm quarks and may be detected
in this decay channel, provided these decays may be disentangled from the ones into bottom quarks.

\subsection{Higgs associated production with charm quarks}

%Similarly to what happens with bottom quarks, the Higgs may be produced in association with charm quarks, as introduced in the previous section. Therefore, searches for the Higgs boson in this channel may provide a further possible constraint on large values of $|\kappa_c|$. \textbf{Account for tagging efficiency of two charm jets, show expected events as function of $\kappa_c$ or find value for which $\sigma_{ccH} = \sigma_{bbH}$; use bounds/expected bounds on bbH with b tagging efficiency to find expected bound on ccH $\to$ $\kappa_c$ (?)}
%Additionally, such large enhancements of $\sigma_{ccH}$ would be expected to affect the measured rates of other processes due to c-jet misidentification. We may quantify these effects by calculating the expected observed rates at the LHC taking into account the mistagging of $c\bar{c}H$ final states into other observed processes.
The $cH$ production mode has also been proposed as a search method for $\kappa_c$.   Because
this channel has a lesser dependence on $\kappa_c$ at very large $|\kappa_c|$ than $c\bar{c}H$, it was not included in the analysis of radiative Higgs decays in Section~\ref{sec:radiative}. However, the $cH$ channel  
has a higher production cross section at small or moderate values of $|\kappa_c|$, preferred by the total Higgs width constraints and precision electroweak measurements analyzed in Section~\ref{sec:precmeas}.
A previous study of this channel~\cite{Brivio:2015fxa} shows that a high luminosity LHC, with 3000~fb$^{-1}$ integrated luminosity at ATLAS and CMS, should be able to probe values of $\kappa_c < 2.5$ at the 95\% C.L. This study leaves all other $\kappa$'s fixed to the SM expectation, varying only $\kappa_c$, and therefore we should reanalyze it taking into account the rise of the $\kappa_i$ along the flat direction. 

The $cH$ production process involves three diagrams at leading order: s-channel and t-channel diagrams with a $c$ propagator and a $c\bar{c}H$ vertex, and an s-channel diagram with a gluon propagator and a $ggH$ vertex. Since the diagram with the $ggH$ vertex is dominant for SM values
of the Higgs couplings, we expect that following the flat direction would further enhance the $cH$ production beyond the values found in \cite{Brivio:2015fxa}. However, this also further enhances the background processes $p p \to gH$ and $p p \to bH$ in addition to the $p p \to c\bar{c}H$ background.

We use MadGraph at leading order in a specialized model file, which includes an effective $ggH$ vertex, to calculate the production rates. We vary the values of $\kappa_c$ and increase $\kappa_g$ and $\kappa_b$ proportionally according to Eq. (\ref{eq:corr}) to obtain the production cross section for each process. Using a charm tagging efficiency of 30\%, a $c\bar{c}H$ mistag rate as $cH$ of 5\%, and $b$ and $g$ mistag rates of 20\% and 1\%, respectively \cite{Aaboud:2018fhh}, we obtain the expected number of events for $\sigma(p p \to X H) \times BR(H \to \gamma\gamma)$ for 3 ab$^{-1}$ integrated luminosity. Although $\sigma(p p \to g H) \gg \sigma(p p \to b H)$, the larger $b$ mistag rate leads to similar background contributions from the two processes. The $c\bar{c}H$ background has a stronger dependence on $\kappa_c$ and therefore contributes an increasing fraction of the background for larger $\kappa_c$. The results are shown in Figure \ref{cH}. 

\begin{figure}[H]
	\centering
	\includegraphics[width=0.7\textwidth]{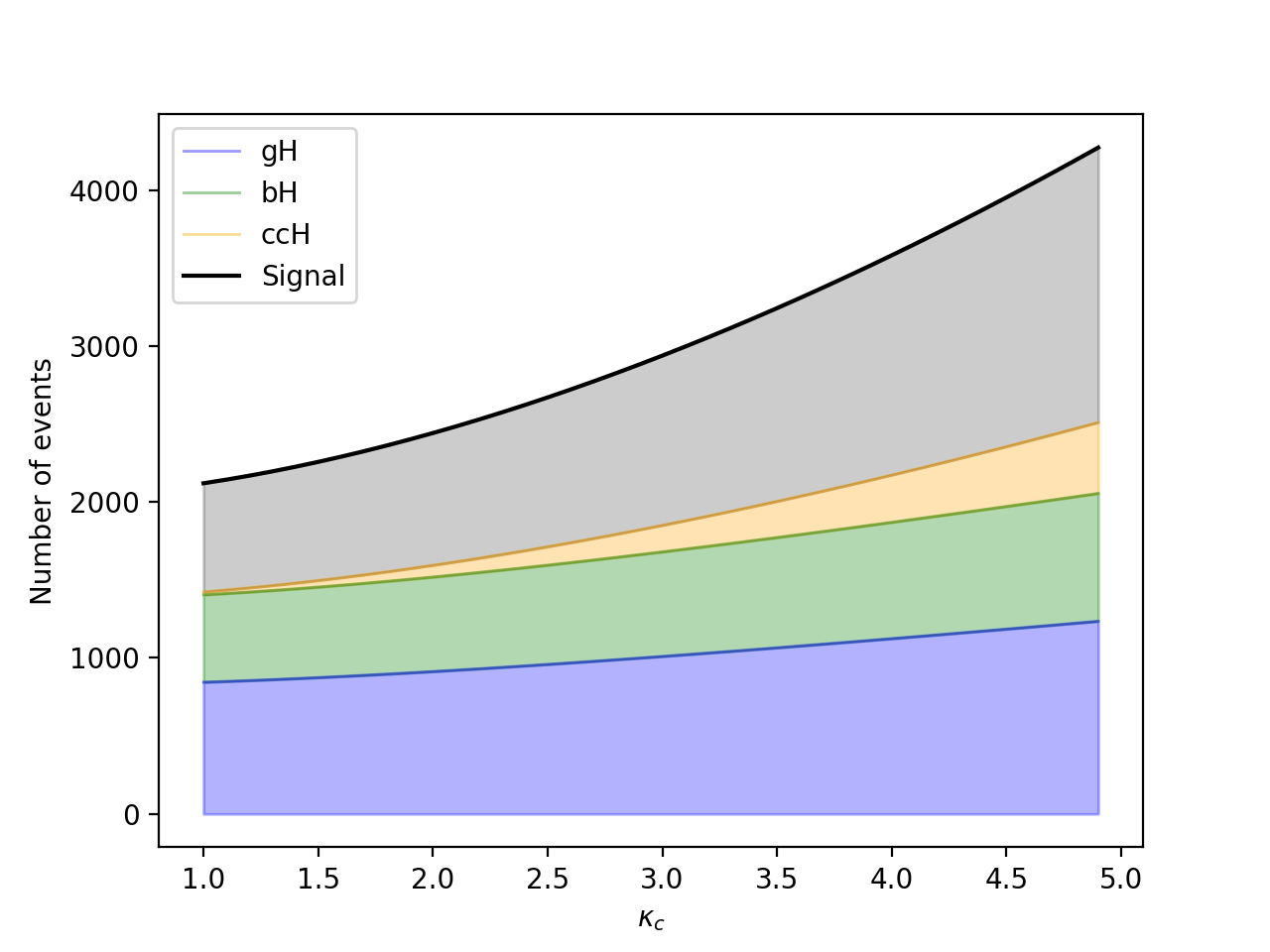}
	\caption{The expected number of background and signal events for $cH$ production at the HL-LHC with 3 ab$^{-1}$ integrated luminosity.}
	\label{cH}
\end{figure}

The $cH$ process includes dependence on both the $\kappa_c$ enhancement and the $\kappa_g$ enhancement along the flat direction. It therefore increases more quickly with $\kappa_c$ than the background processes, which each depend on only one of these enhancements; in particular, the dominant backgrounds of $pp \to bH,gH$ depend only on the flat direction enhancements of $\kappa_{b},\kappa_{g}$. We show the number of signal and background events, along with their ratio, for a range of $\kappa_c$ values in Table \ref{tab:cHevents}.

\begin{table}[H]
	\centering
	\begin{tabularx}{0.8\textwidth}{|X||X|X|X|X|X|X|X|X|X|}
		\hline 
		$\kappa_c$ & 1.0 & 1.5 & 2.0 & 2.5 & 3.0 & 3.5 & 4.0 & 4.5 & 5.0 \\ \hline \hline
		$S$ & 687 & 758 & 840 & 961 & 1085 & 1230 & 1408 & 1598 & 1822 \\ \hline
		$B$ & 1425 & 1498 & 1595 & 1714 & 1852 & 2005 & 2174 & 2356 & 2551 \\ \hline
		$S/B$ & 0.33 & 0.34 & 0.35 & 0.36 & 0.37 & 0.38 & 0.39 & 0.40 & 0.42 \\ \hline
	\end{tabularx}
	\caption{The number of signal events, number of background events, and signal to background ratio for values of $\kappa_c$ between 1 and 5. Due to the increase in $\kappa_g,\kappa_b$ along the flat direction, the background increases in addition to the signal.}
	\label{tab:cHevents}
\end{table}

Since variations in $\sigma_{cH}$ depend weakly on $\kappa_c$ alone along the flat direction, it would be very difficult to identify the precise value of $\kappa_c$ from a measurement of $N=S+B$. However, we may use these signal and background rates to estimate the sensitivity to $\kappa_c$ following a similar analysis to the one in Ref.~\cite{Brivio:2015fxa}. Assuming the true value of $\kappa_c$ is 1, we find the expected 1$\sigma$ and 2$\sigma$ upper bounds on $\kappa_c$ from this process by identifying the value of $\kappa_c$ for which $N(\kappa_c) - N(1) = 1\sigma,2\sigma$. We take the statistical uncertainty to be $\Delta N^{stat} (\kappa_c) = \sqrt{S(\kappa_c) + B(\kappa_c)}$ and the theoretical uncertainty in the signal and background, which we have calculated at LO, to be 20\%. Because our background is now also being estimated for varying $\kappa_c$ using MadGraph5, we examine two cases for the uncertainty in the background. In the first case, we apply no uncertainty to the number of background events. In the second case, we apply a 20\% uncertainty to the number of background events $B(\kappa_c)$ in addition to the number of signal events. We find $\Delta N^{tot}$ by adding the statistical and theoretical uncertainties in quadrature. Let us stress that this analysis assumes that the dominant uncertainties are the statistical and theoretical ones and ignores the possible impact of systematic and experimental uncertainties. The sensitivity on $\kappa_c$ depends strongly on these assumptions, and may become weaker after a realistic experimental analysis of this process is performed.

We take  $\Delta N^{tot} = \sigma$ to parametrize the number of standard deviations of $N(\kappa_c) - N(1) = n\sigma$ for the two uncertainty cases. The value of $n$  is plotted versus $\kappa_c$ in Fig. \ref{cH_sigmas}. We find a 1$\sigma$ (2$\sigma$) deviations for
\begin{equation}
	|\kappa_c| < 1.6 \; (2.1)
\end{equation}
in the first case, and
\begin{equation}
	|\kappa_c| < 2.5 \; (4.0)
\end{equation}
in the second case.   In the first case the increase of the expected sensitivity relative to~\cite{Brivio:2015fxa},  in which no uncertainty was applied to the background estimates,
arises from the enhancement of the background events in addition to the signal events. In the second case, we find approximately the same expected sensitivity as in Ref.~\cite{Brivio:2015fxa}. 

Although the best-fit $\kappa$ values for low values of $\kappa_c$ tend to follow the flat direction, we note that taking SM-like values for the other couplings can still retain some level of consistency with LHC results for this range of $\kappa_c$; therefore, our results do not invalidate the analysis of Ref.~\cite{Brivio:2015fxa} but show the variation of the LHC sensitivity for slightly larger  values of $\kappa_g$, for which 
an improvement of the fit to the Higgs precision measurement data is obtained.

\begin{figure}[H]
	\centering
	\subfloat[][]{ \includegraphics[width=0.45\textwidth]{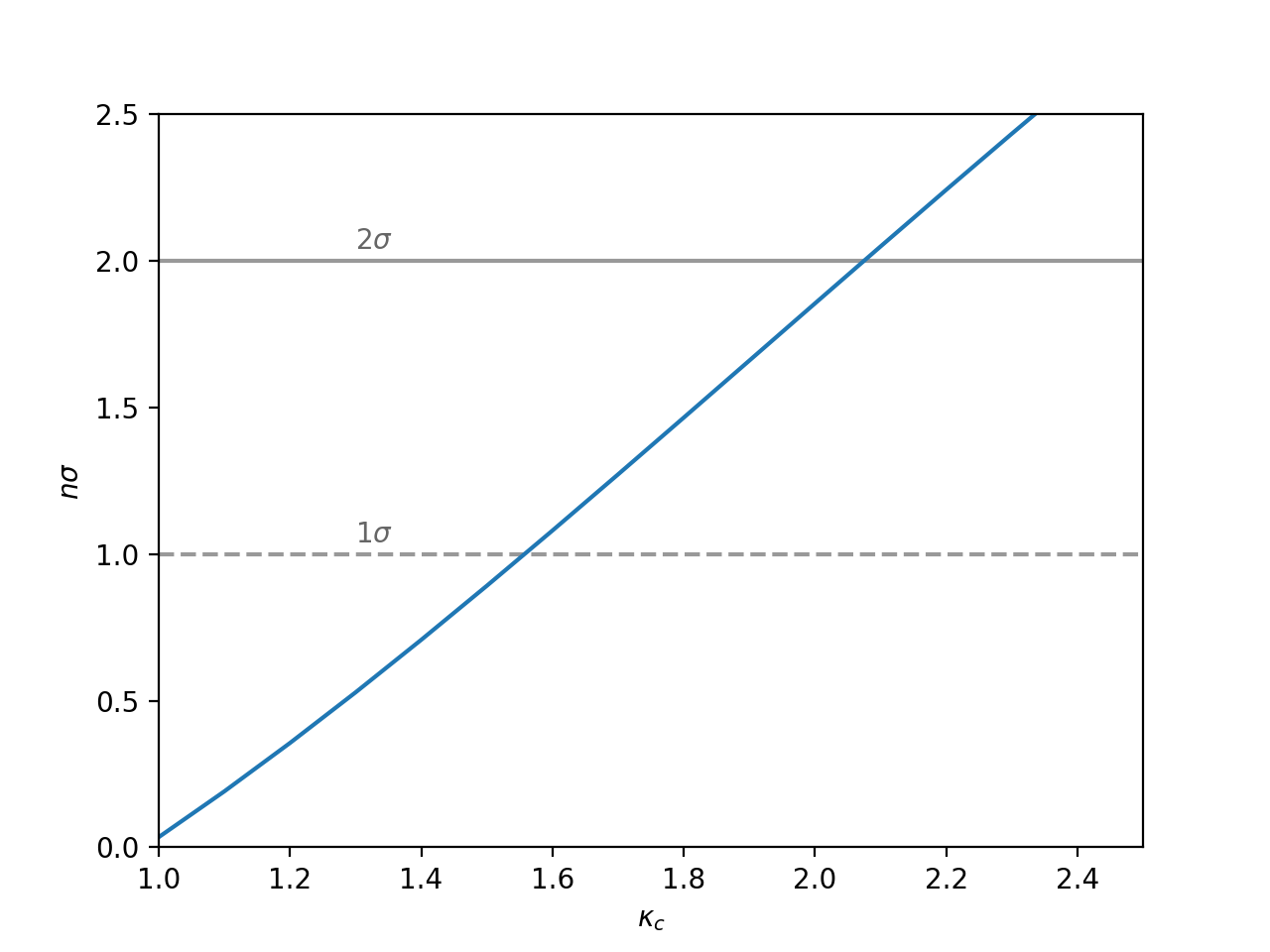} }
	\subfloat[][]{ \includegraphics[width=0.45\textwidth]{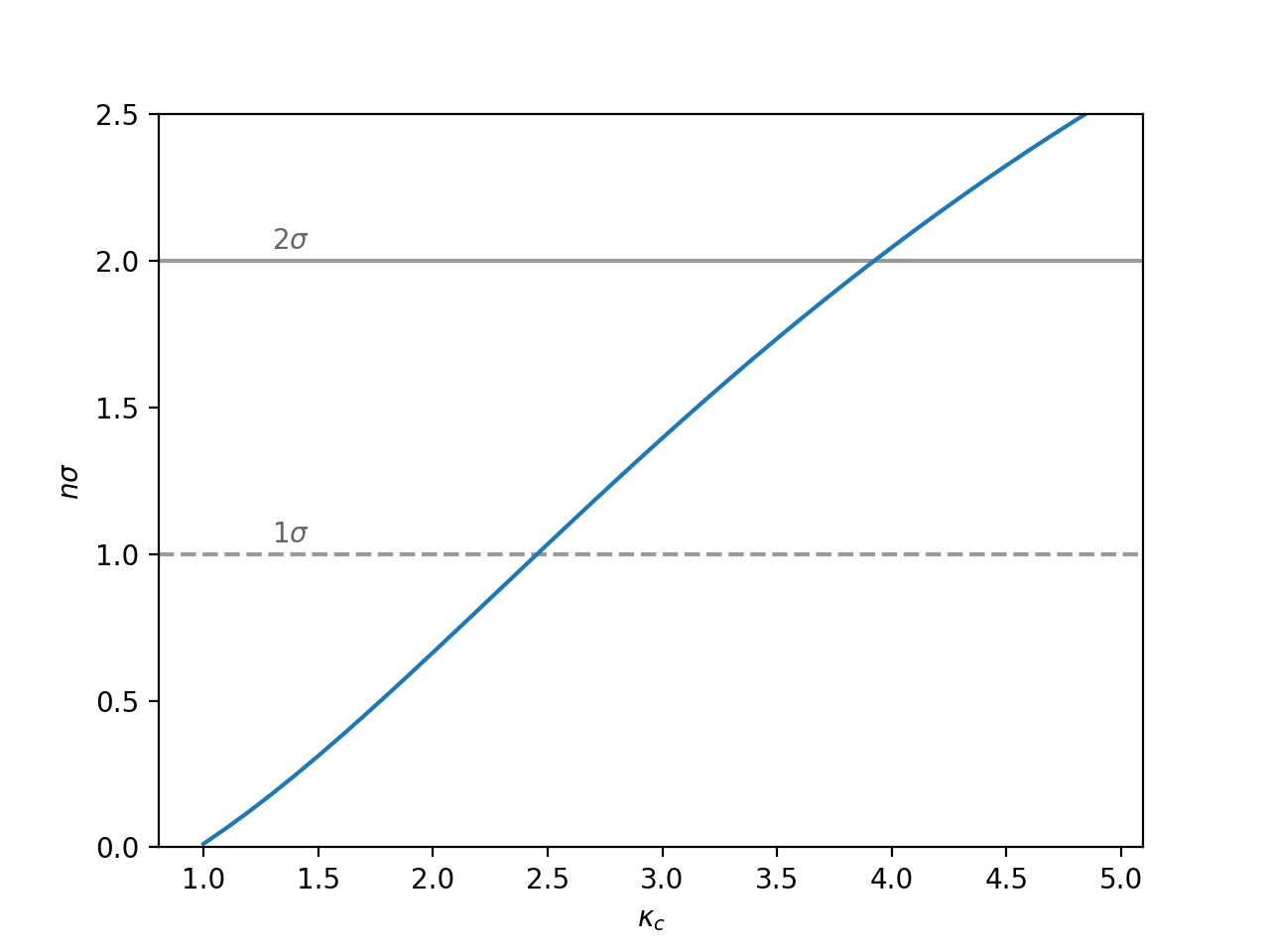} }
	\caption{Number of standard deviations of $N(\kappa_c)$ from $N(1)$, as a function of $\kappa_c$. The dashed (solid) grey lines indicate the 1$\sigma$ (2$\sigma$) bounds. The two cases represent (left) no uncertainty in background and (right) 20\% uncertainty on the number of background events.}
	\label{cH_sigmas}
\end{figure}

\subsection{Higgs decay into charm-quark pairs}

\subsubsection{Direct Searches}

Searches have been performed for $ZH \to l^{+} l^{-} c\bar{c}$ with 36.1 fb$^{-1}$ integrated luminosity, with ATLAS publishing an upper bound of $\sigma(pp \to ZH) \times B(H \to cc) < 2.7$ pb at 95\% CL \cite{Aaboud:2018fhh}. This corresponds to about 110 times the SM rate. Thus we require that $\kappa_Z^2 \kappa_c^2 / \kappa_H^2 \lesssim 110$; moving along the flat direction, one reaches this limit at a value of $|\kappa_c| = 20.9$, which is a far weaker bound than the one provided by the total width constraints. However, HL-LHC studies from ATLAS \cite{ATLAS:ZhUpgrade} have found an expected upper bound of $\mu_{ZH\to c\bar{c}} < 6.3$ at 95\% CL with an integrated luminosity of 3 ab$^{-1}$. Unconstrained fits of the rate measurements remain within this limit for $|\kappa_c| \lesssim 2.7$; this channel may therefore provide a bound of similar magnitude to those from constrained-fit bounds at the HL-LHC.

The $ZH \to l^{+} l^{-} c\bar{c}$ limit obtained in the ATLAS HL-LHC study uses a tighter charm tagging working point than the working point employed in Run 2, thereby reducing the background contribution from processes such as $ZH \to Z b\bar{b}$. In particular, the tagging efficiency for c-jet, and mis-tagging rates for b-jet, and light-flavor jets are 18\%, 5\%, and 0.5\%, respectively, for the HL-LHC study, while these values are 41\%, 25\%, and 5\% for the Run 2 analysis. This stricter working point takes advantage of the higher expected signal yield at the HL-LHC to provide a 7\% additional improvement on the limit relative to Run 2. However, charm tagging algorithms are currently being improved, in part through the use of deep neural networks. For example, CMS deep tagging algorithms have achieved a 24\% tagging efficiency with 1\% b-jet and 0.2\% light jet mis-tagging rates \cite{CMStwiki}. This algorithm therefore has a 6\% improvement in efficiency over the HL-LHC study working point along with a factor 5 improvement in the b-jet mis-tag rate. The use of new tagging algorithms could therefore further improve the limit obtained at the HL-LHC. 

\subsubsection{Indirect Searches}

The $H \to c\bar{c}$ decay can also be examined in the context of $H \to b\bar{b}$ decays to place a bound on $\kappa_c$ using current data \cite{Perez:2015aoa,Perez:2015lra}. We examine the effect of $c\bar{c}$ mistagging as $b\bar{b}$ on the observed $H \to b\bar{b}$ rates. This results in $\kappa_c$ being a factor in the numerator of $\mu_{i,b\bar{b}}$, thereby limiting the flat direction described by Eq.~(\ref{eq:corr}) for large values of 
$\kappa_c$. 
We include the $c\bar{c}$ contributions to $b\bar{b}$ rates by
\begin{equation}
	\mu_{i,b\bar{b}} = \kappa_{i}^2 \frac{ \kappa_b^2 + \kappa_c^2 (BR_{c\bar{c}} \epsilon_{c}^2 / BR_{b\bar{b}} \epsilon_b^2) } {\kappa_H^2}
\end{equation}
where $\epsilon_{c}$ is the mistag rate of $c$-jets as $b$-jets and $\epsilon_b$ is the tagging efficiency of $b$-jets and we have defined $\mu_{i,b\bar{b}}$ as the observed rate normalized to the uncontaminated SM rate. Our analysis of this bound differs from that by Perez et.~al., Ref.~\cite{Perez:2015lra}, in two primary ways. Firstly, we include this altered expression for $\mu_{i,b\bar{b}}$ in our fit to all of the LHC observed rates listed in Table \ref{tab:channels}, thereby removing the `flat direction' for $\mu_{i,b\bar{b}}$ along $\kappa_b = \kappa_c$ encountered in~\cite{Perez:2015aoa}, which examines only $H \to b\bar{b}$ processes. We therefore do not need to employ multiple tagging points to obtain a bound for $\kappa_c$, since for sizable values of $\kappa_c$, raising $\kappa_b$ and $\kappa_c$ together will spoil the fit to other observables. Consequently, we allow variations in the other $\kappa$'s, which approximately follow the flat direction described by Eq.~(\ref{eq:corr}). Because of this, $\kappa_b$ and $\kappa_c$ may have greater variations that those found in Refs. \cite{Perez:2015aoa,Perez:2015lra} while remaining consistent with observed $b\bar{b}$ (and all other) Higgs rates. We therefore expect to find weaker bounds in our analysis of this potential bound.

We employ the ATLAS working point of $\epsilon_b = 0.70$, $\epsilon_c = 0.20$ and the CMS working point of $\epsilon_b = 0.78$, $\epsilon_c = 0.27$. To obtain a bound, we perform a fit to the Higgs rate measurements and place a limit on $\Delta \chi^2$. Following this analysis, the ATLAS and CMS tagging efficiencies provide bounds of $|\kappa_c| \lesssim 23$ and $|\kappa_c| \lesssim 16$, respectively. Using the HL-LHC expected uncertainties~\cite{ATLAS_mu,CMS:2018qgz} along with best-fit rates of $\mu=1.0$, this approach places bounds of $|\kappa_c| \lesssim 8.7$ and $|\kappa_c| \lesssim 6.5$, respectively.  

\subsection{Asymmetry in $W^{+}H$ and $W^{-}H$ production}

The measurement of asymmetry in $\sigma( p p \to W^{+} H )$ and $\sigma( p p \to W^{-} H )$ production has also been proposed as a channel through which one can place limits on $\kappa_c$ \cite{Yu:2016rvv}. The relevant diagrams for this process are shown in Fig.~\ref{W_diagrams}. The SM asymmetry is driven by the Higgs-Strahlung processes; in the Higgs-Strahlung diagrams, the difference in $W^{+}$ and $W^{-}$ production arises from the asymmetry of $u\bar{d}$ and $\bar{u}d$ in the proton PDF. The charm Yukawa appears in diagrams with $s\bar{c}$ and $\bar{s}c$ initial states, which are symmetric in the proton PDF. Therefore, when the charm Yukawa is increased significantly, the symmetric $s\bar{c}$/$\bar{s}c$ diagrams reduce the asymmetry with respect to the SM expected value. The $W^{\pm}H$ production asymmetry therefore decreases with large $\kappa_c$.
One can therefore use the sensitivity of this asymmetry on $\kappa_c$ to get bounds on the charm coupling~\cite{Yu:2016rvv}. 

\begin{figure}[H]
	\centering
	\includegraphics[width=0.45\textwidth]{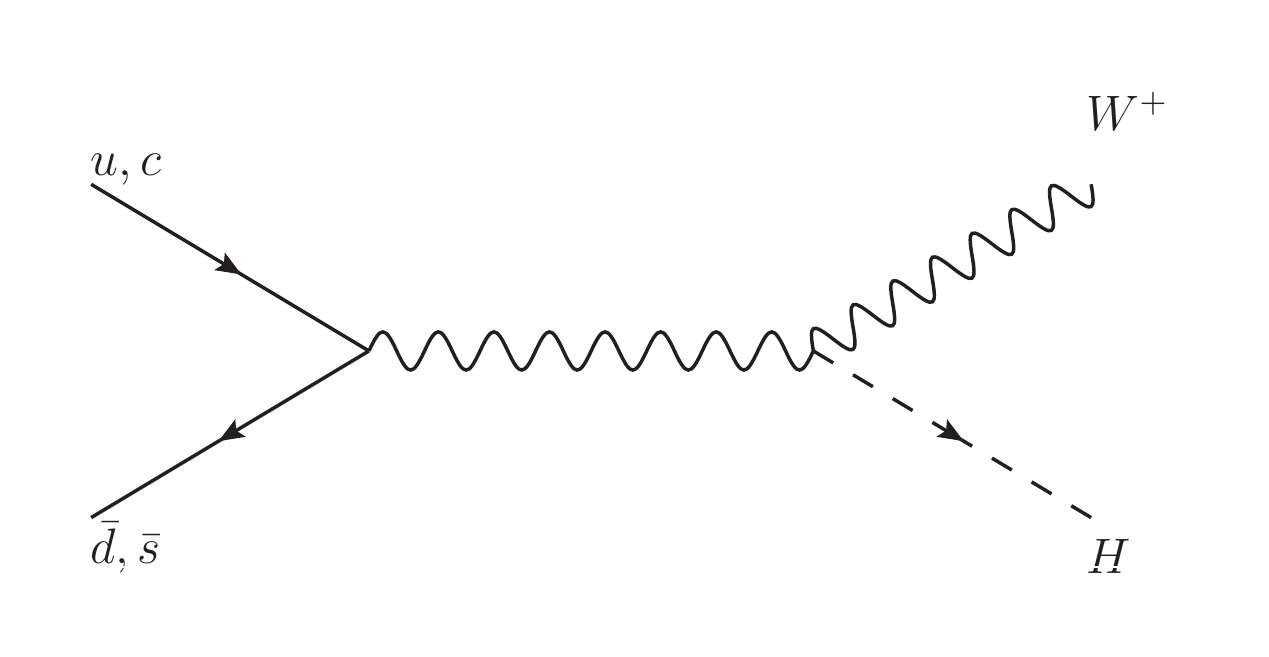}
	\includegraphics[width=0.45\textwidth]{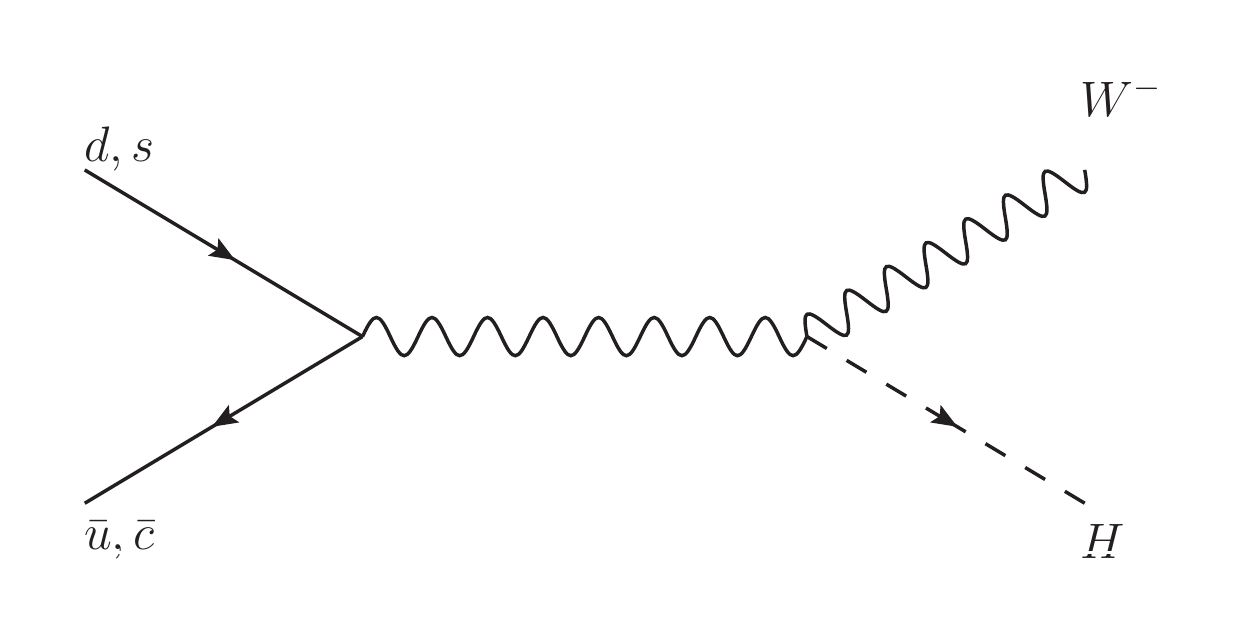} \\
	\includegraphics[width=0.45\textwidth]{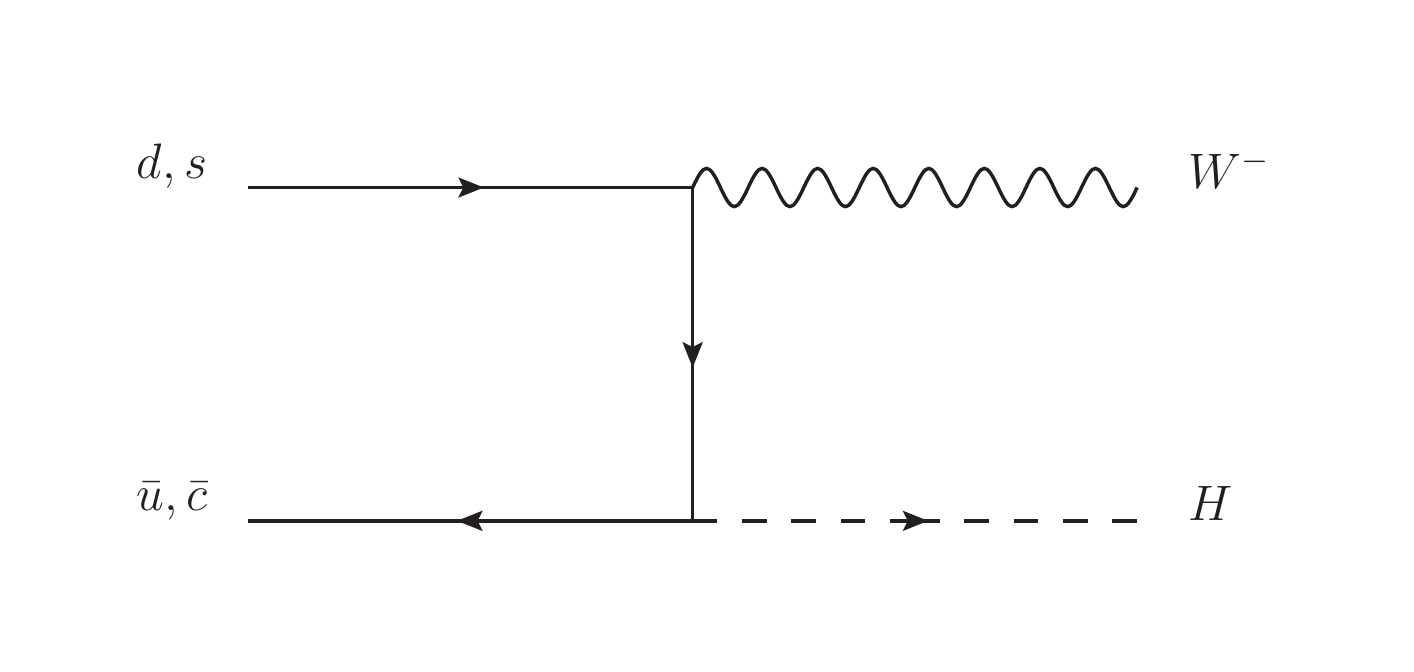}
	\includegraphics[width=0.45\textwidth]{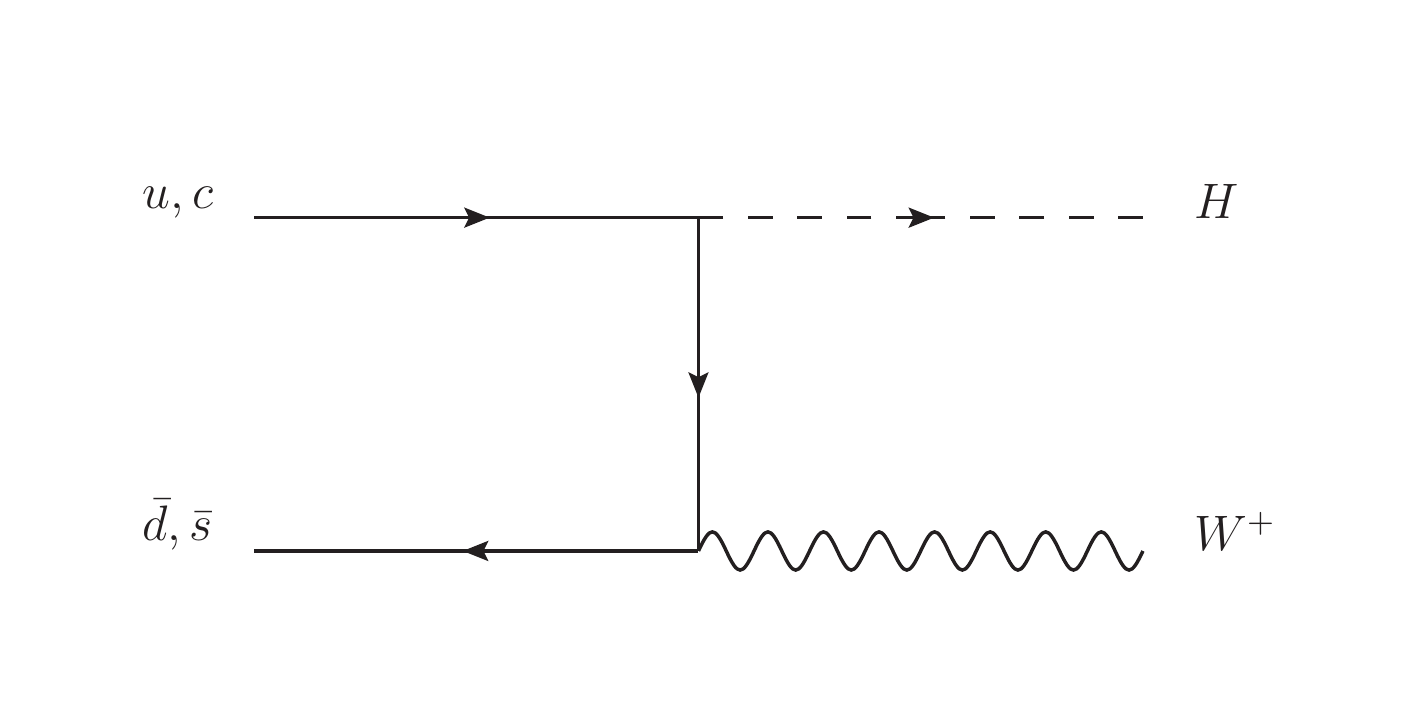}
	\caption{Diagrams for the two relevant types of $W^{\pm}H$ production processes at leading order. The top row shows the Higgstrahlung processes, which are dominant in the SM, while the bottom row shows the diagrams proportional to the charm Yukawa.}
	\label{W_diagrams}
\end{figure}

Given the relative contributions of the two types of diagrams, however, we note that enhancements of $\kappa_W$ alongside enhancements of $\kappa_c$ will reduce the symmetrizing effect of increasing $\kappa_c$. In order to examine this quantitatively, we use MadGraph5 to calculate the LO cross sections at 14 TeV for $W^{+}H$ and $W^{-}H$ production along the flat direction. Figure \ref{fig:Wasym} shows the results of this analysis. We plot the percent asymmetry of the production modes, quantified as
\begin{equation}
	A_{W^{\pm}} = \frac{ \sigma_{W^{+}H} - \sigma_{W^{-}H} } {\sigma_{W^{+}H} + \sigma_{W^{-}H} },
\end{equation}
as a function of $\kappa_c$ along the flat direction, as well as for $\kappa_X=1$ with $X\neq c$.

We find that the asymmetry is reduced to less than 0.02 up to $\kappa_c=100$.  Using MadGraph5 and detector simulations, Ref. \cite{Yu:2016rvv} found that the uncertainty in the asymmetry may be reduced to approximately 0.004 with 3 ab$^{-1}$ integrated luminosity. In this case, the $W^{\pm}$ asymmetry would be able to place a limit of $|\kappa_c| \lesssim 30$ along the flat direction. This still provides a weaker bound than other proposed methods by approximately an order of magnitude, and we therefore conclude that if one requires consistency with LHC precision Higgs measurements, the $W^{\pm}H$ asymmetry does not provide a sensitive probe of $\kappa_c$.

\begin{figure}[H]
	\centering
	\includegraphics[width=0.7\textwidth]{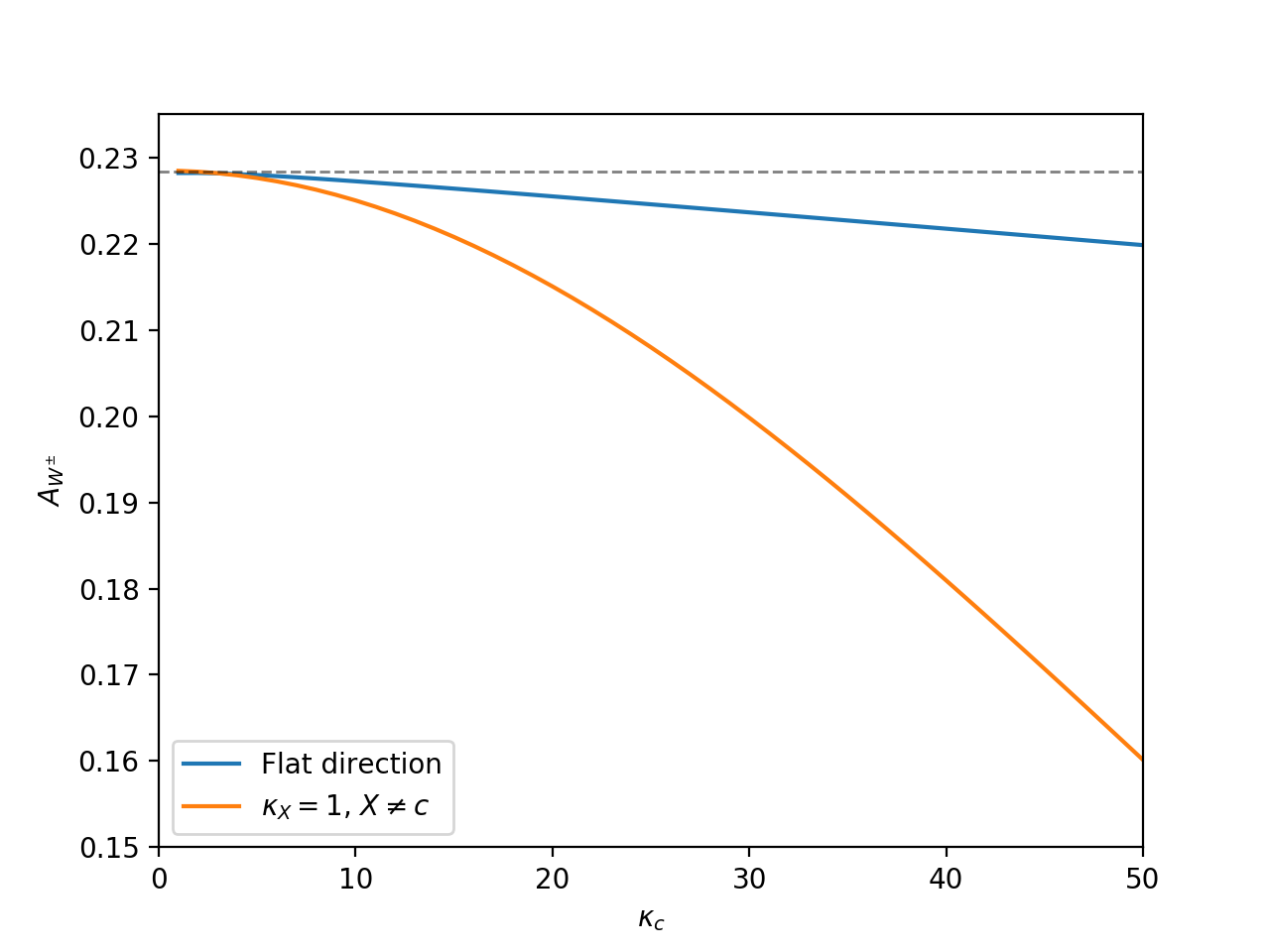}
	\caption{Plot of the percent asymmtry in $W^{\pm}H$ production versus $\kappa_c$, for the flat direction and for SM-like $\kappa_X$, $X \neq c$. While large $\kappa_c$ significantly reduces the asymmetry in the second case, the enhancement of $\kappa_W$ alongside $\kappa_c$ in the flat direction reduces the relative effect of the symmetrizing $\kappa_c$-proportional contributions.}
	\label{fig:Wasym}
\end{figure}

\subsection{Differential cross sections}

The distribution of the Higgs production differential cross section as a function of transverse momentum has also been proposed as a probe of $\kappa_c$~\cite{Bishara:2016jga, Soreq:2016rae, Bonner:2016sdg} and has been examined for 35.9 fb$^{-1}$ of data by CMS~\cite{Sirunyan:2018sgc}. This method of bounding $\kappa_c$ may provide an interesting complementary bound to those from the fit to precision rate measurements, as the flat direction along which the rates remain constant may not reproduce the expected SM cross section distribution as a function of transverse momentum. The CMS study examines the $H \to \gamma \gamma$ and $H \to ZZ$ decay channels, as well as their combination, and identifies bounds by varying $\kappa_b$ and $\kappa_c$ and examining two cases: the first in which the branching fractions are dependent on $\kappa_{b,c}$, and the second in which they are independent. In the dependent case, they quote a bound of $-4.9 < \kappa_c < 4.8$, while in the independent case the bound is $-33 < \kappa_c < 38$. The uncertainties in the cross section distribution, which are on the order of 10-20\%, are currently dominated by statistical uncertainty, while the systematic uncertainty is on the order of about 5\%. The bounds quoted above would therefore be expected to improve with more data.

However, we again note that varying $\kappa_{b,c}$ to values as large as 5 would significantly affect the other observed channels, and that therefore the flat direction is necessary to ensure consistency with the current Higgs observations. It is likely that varying the other couplings along the flat direction will affect the bound in this case. In particular, the variation of $\kappa_t$ in addition to $\kappa_b$ and $\kappa_c$ should affect the expected distribution and would likely weaken the identified bounds, while the branching fractions would vary less dramatically with increases in $\kappa_{b,c}$. One might expect that along the flat direction the bounds will be similar to the one found in the unconstrained case. A study of this bound with the addition of the flat direction is necessary to provide a bound on $\kappa_c$ that is consistent with the other LHC measurements.

Ref.~\cite{Bishara:2016jga} has predicted the possible HL-LHC bounds from the differential cross section distributions. Assuming a theory uncertainty of 2.5\% and systematic uncertainty of 1.5\%, they find a 95\% CL bound of $\kappa_c \in [-0.6, 3.0]$. However, we emphasize that these bounds do not take into account the rate measurements and the flat direction, and also assume significant improvements in the theoretical and systematic uncertainties.

\section{Conclusions}
\label{sec:Conc}

After the Higgs discovery, one of the main goals of the High Energy Program is the detailed study of
its properties. In particular, the measurement of the Higgs couplings to SM bosons and fermions is
of crucial importance. Most of the  Higgs production and decay processes measured at the LHC
are sensitive to the gauge bosons and third generation quark and lepton Yukawa couplings and therefore,
considering only variations of these couplings, they are being determined within an accuracy of the order of tens of percent.

The first and second generation quark and lepton couplings are, however, not yet determined. In particular,
the Yukawa coupling of the charm quark, characterized by $\kappa_c$  in the $\kappa$ framework, is only weakly
constrained.  In this work we updated the bounds on $\kappa_c$, paying particular attention to the consistency
with the LHC Higgs precision measurements.  In this sense, we discussed the existence of particular correlations
between the charm coupling and the gauge boson and third generation couplings that allow consistency
with the measured Higgs process rates, even for large deviations of $\kappa_c$. 

Due to the existence of these correlations, a bound on $\kappa_c$ may only be obtained by imposing additional
constraints. These are provided by bounds on the Higgs width, precision measurements, and $|\kappa_V| \leq 1$, 
 leading to a 95\% CL bound on $|\kappa_c| < 7.5$,~4.9, and 2.7, respectively.  The Higgs width and $|\kappa_V| \leq 1$ bounds
may be improved at higher luminosities to values of order $|\kappa_c| \lesssim 3.0$ and 2.1, respectively. 

We also analyzed radiative
decays of the Higgs into quarkonium states, explaining the relevance of the flat direction and the variations of
the Higgs width and the production rate. No competitive bound on $\kappa_c$ from LHC data may be obtained,
even at high luminosities. 

Finally, we studied Higgs processes induced by the charm-quark Yukawa coupling. These include both Higgs
production in association with charm quarks as well as possible decays of the Higgs into charm states. While
currently all these searches cannot provide a competitive bound on $\kappa_c$, the possible improvements
in charm tagging at higher luminosities may lead to a sensitivity that is similar to the one obtained from precision
Higgs measurements, namely $|\kappa_c| \lesssim 2$ and 2.7 in the $cH$ and $ZH, H\to c\bar{c}$ channels, respectively. The effect of $\kappa_c$ on the differential Higgs production cross section may also provide a competitive bound, but it will demand an improvement in the current theoretical and systematic uncertainties. Moreover, a careful examination of this bound, taking into account all observed Higgs rates, should be performed.

\section{Acknowledgements}

We thank Javier Duarte, Florian Goertz, Gino Isidori and Konstantinos Nikolopoulos for useful discussions. Work at ANL is supported in part by the U.S. Department of Energy under Contract No. DE-AC02-06CH11357. The work of C.W. and N.C. at EFI is supported by the U.S. Department of Energy under Contract No. DE-FG02-13ER41958.  V.W. is supported by the University of Chicago Physics Department.


\begin{thebibliography}{99}

%\cite{Tanabashi:2018oca}
\bibitem{Tanabashi:2018oca} 
  M.~Tanabashi {\it et al.} [Particle Data Group],
  %``Review of Particle Physics,''
  Phys.\ Rev.\ D {\bf 98}, no. 3, 030001 (2018).
  doi:10.1103/PhysRevD.98.030001
  %%CITATION = doi:10.1103/PhysRevD.98.030001;%%
  %1392 citations counted in INSPIRE as of 13 Apr 2019

%\cite{Group:2012gb}
\bibitem{Group:2012gb} 
  T.~E.~W.~Group [CDF and D0 Collaborations],
  %``2012 Update of the Combination of CDF and D0 Results for the Mass of the W Boson,''
  arXiv:1204.0042 [hep-ex].
  %%CITATION = ARXIV:1204.0042;%%
  %158 citations counted in INSPIRE as of 20 Mar 2019
  
  %\cite{ALEPH:2005ab}
\bibitem{ALEPH:2005ab} 
  S.~Schael {\it et al.} [ALEPH and DELPHI and L3 and OPAL and SLD Collaborations and LEP Electroweak Working Group and SLD Electroweak Group and SLD Heavy Flavour Group],
  %``Precision electroweak measurements on the $Z$ resonance,''
  Phys.\ Rept.\  {\bf 427}, 257 (2006)
  %doi:10.1016/j.physrep.2005.12.006
  [hep-ex/0509008].
  %%CITATION = doi:10.1016/j.physrep.2005.12.006;%%
  %1967 citations counted in INSPIRE as of 20 Mar 2019
  
  %\cite{LEP:2003aa}
\bibitem{LEP:2003aa} 
  t.~S.~Electroweak [LEP and ALEPH and DELPHI and L3 and OPAL Collaborations and LEP Electroweak Working Group and SLD Electroweak Group and SLD Heavy Flavor Group],
  %``A Combination of preliminary electroweak measurements and constraints on the standard model,''
  hep-ex/0312023.
  %%CITATION = HEP-EX/0312023;%%
  %310 citations counted in INSPIRE as of 20 Mar 2019
  
  %\cite{Schael:2013ita}
\bibitem{Schael:2013ita} 
  S.~Schael {\it et al.} [ALEPH and DELPHI and L3 and OPAL and LEP Electroweak Collaborations],
  %``Electroweak Measurements in Electron-Positron Collisions at W-Boson-Pair Energies at LEP,''
  Phys.\ Rept.\  {\bf 532}, 119 (2013)
 % doi:10.1016/j.physrep.2013.07.004
  [arXiv:1302.3415 [hep-ex]].
  %%CITATION = doi:10.1016/j.physrep.2013.07.004;%%
  %374 citations counted in INSPIRE as of 20 Mar 2019

%\cite{ATLAS:2018doi}
\bibitem{ATLAS:2018doi} 
  The ATLAS collaboration [ATLAS Collaboration],
  %``Combined measurements of Higgs boson production and decay using up to 80 fb$^{-1}$ of proton--proton collision data at $\sqrt{s}=$ 13 TeV collected with the ATLAS experiment,''
  ATLAS-CONF-2018-031.
  %%CITATION = ATLAS-CONF-2018-031;%%
  %28 citations counted in INSPIRE as of 08 Jan 2019
  
  %\cite{Aaboud:2018zhk}
\bibitem{Aaboud:2018zhk} 
  M.~Aaboud {\it et al.} [ATLAS Collaboration],
  %``Observation of $H \rightarrow b\bar{b}$ decays and $VH$ production with the ATLAS detector,''
  Phys.\ Lett.\ B {\bf 786}, 59 (2018)
 % doi:10.1016/j.physletb.2018.09.013
  [arXiv:1808.08238 [hep-ex]].
  %%CITATION = doi:10.1016/j.physletb.2018.09.013;%%
  %29 citations counted in INSPIRE as of 09 Jan 2019
  
  %\cite{Sirunyan:2018koj}
\bibitem{Sirunyan:2018koj} 
  A.~M.~Sirunyan {\it et al.} [CMS Collaboration],
  %``Combined measurements of Higgs boson couplings in proton-proton collisions at $\sqrt{s}=$ 13 TeV,''
  %Submitted to: Eur.Phys.J.
  [arXiv:1809.10733 [hep-ex]].
  %%CITATION = ARXIV:1809.10733;%%
  %36 citations counted in INSPIRE as of 25 Feb 2019
  
  %\cite{LHCHiggsCrossSectionWorkingGroup:2012nn}
\bibitem{LHCHiggsCrossSectionWorkingGroup:2012nn} 
  A.~David {\it et al.} [LHC Higgs Cross Section Working Group],
  %``LHC HXSWG interim recommendations to explore the coupling structure of a Higgs-like particle,''
  arXiv:1209.0040 [hep-ph].
  %%CITATION = ARXIV:1209.0040;%%
  %224 citations counted in INSPIRE as of 30 May 2019
  
  %\cite{Heinemeyer:2013tqa}
\bibitem{Heinemeyer:2013tqa} 
  S.~Heinemeyer {\it et al.} [LHC Higgs Cross Section Working Group],
  %``Handbook of LHC Higgs Cross Sections: 3. Higgs Properties,''
  doi:10.5170/CERN-2013-004
  arXiv:1307.1347 [hep-ph].
  %%CITATION = doi:10.5170/CERN-2013-004;%%
  %1206 citations counted in INSPIRE as of 30 May 2019
  
%\cite{Gupta:2014rxa}
\bibitem{Gupta:2014rxa} 
  R.~S.~Gupta, A.~Pomarol and F.~Riva,
  %``BSM Primary Effects,''
  Phys.\ Rev.\ D {\bf 91}, no. 3, 035001 (2015)
  doi:10.1103/PhysRevD.91.035001
  [arXiv:1405.0181 [hep-ph]].
  %%CITATION = doi:10.1103/PhysRevD.91.035001;%%
  %97 citations counted in INSPIRE as of 18 Apr 2019

%\cite{Gupta:2014rxa}
%\cite{Contino:2013kra}
\bibitem{Contino:2013kra} 
  R.~Contino, M.~Ghezzi, C.~Grojean, M.~Muhlleitner and M.~Spira,
  %``Effective Lagrangian for a light Higgs-like scalar,''
  JHEP {\bf 1307}, 035 (2013)
  doi:10.1007/JHEP07(2013)035
  [arXiv:1303.3876 [hep-ph]].
  %%CITATION = doi:10.1007/JHEP07(2013)035;%%
  %315 citations counted in INSPIRE as of 19 Apr 2019
    
 %\cite{Gupta:2014rxa}
%\cite{Contino:2013kra}
  %\cite{Falkowski:2015fla}
\bibitem{Falkowski:2015fla} 
  A.~Falkowski,
  %``Effective field theory approach to LHC Higgs data,''
  Pramana {\bf 87}, no. 3, 39 (2016)
  doi:10.1007/s12043-016-1251-5
  [arXiv:1505.00046 [hep-ph]].
  %%CITATION = doi:10.1007/s12043-016-1251-5;%%
  %68 citations counted in INSPIRE as of 18 Apr 2019
  
  
   %\cite{Gupta:2014rxa}
%\cite{Contino:2013kra}
  %\cite{Falkowski:2015fla}
  %\cite{Buchalla:1997kz}
%\cite{deFlorian:2016spz}
\bibitem{deFlorian:2016spz} 
  D.~de Florian {\it et al.} [LHC Higgs Cross Section Working Group],
  %``Handbook of LHC Higgs Cross Sections: 4. Deciphering the Nature of the Higgs Sector,''
  doi:10.23731/CYRM-2017-002
  arXiv:1610.07922 [hep-ph].
  %%CITATION = doi:10.23731/CYRM-2017-002;%%
  %642 citations counted in INSPIRE as of 18 Apr 2019


    %\cite{Zeppenfeld:2000td}
\bibitem{Zeppenfeld:2000td} 
  D.~Zeppenfeld, R.~Kinnunen, A.~Nikitenko and E.~Richter-Was,
  %``Measuring Higgs boson couplings at the CERN LHC,''
  Phys.\ Rev.\ D {\bf 62}, 013009 (2000)
  doi:10.1103/PhysRevD.62.013009
  [hep-ph/0002036].
  %%CITATION = doi:10.1103/PhysRevD.62.013009;%%
  %309 citations counted in INSPIRE as of 15 Apr 2019
  
    %\cite{Zeppenfeld:2000td}
  %\cite{Djouadi:2000gu}
\bibitem{Djouadi:2000gu} 
  A.~Djouadi {\it et al.},
  ``The Higgs working group: Summary report,''
  hep-ph/0002258.
  %%CITATION = HEP-PH/0002258;%%
  %117 citations counted in INSPIRE as of 15 Apr 2019
  %\cite{Duhrssen:2004cv}

\bibitem{Duhrssen:2004cv} 
  M.~Duhrssen, S.~Heinemeyer, H.~Logan, D.~Rainwater, G.~Weiglein and D.~Zeppenfeld,
  %``Extracting Higgs boson couplings from CERN LHC data,''
  Phys.\ Rev.\ D {\bf 70}, 113009 (2004)
  doi:10.1103/PhysRevD.70.113009
  [hep-ph/0406323].
  %%CITATION = doi:10.1103/PhysRevD.70.113009;%%
  %346 citations counted in INSPIRE as of 15 Apr 2019
  
    %\cite{Belanger:2013xza}
\bibitem{Belanger:2013xza} 
  G.~Belanger, B.~Dumont, U.~Ellwanger, J.~F.~Gunion and S.~Kraml,
  %``Global fit to Higgs signal strengths and couplings and implications for extended Higgs sectors,''
  Phys.\ Rev.\ D {\bf 88}, 075008 (2013)
  doi:10.1103/PhysRevD.88.075008
  [arXiv:1306.2941 [hep-ph]].
  %%CITATION = doi:10.1103/PhysRevD.88.075008;%%
  %259 citations counted in INSPIRE as of 13 Apr 2019

\bibitem{PCA_1}
  Karl Pearson F.R.S. (1901) LIII,
  %On lines and planes of closest fit to systems of points in space, 
  The London, Edinburgh, and Dublin Philosophical Magazine and Journal of Science, 2:11, 559-572, DOI: 10.1080/14786440109462720
  
\bibitem{PCA_2}
   Hotelling, H. (1933), 
   %Analysis of a complex of statistical variables into principal components. Journal of Educational Psychology, 24(6), 417-441.
   http://dx.doi.org/10.1037/h0071325

%\cite{Aaboud:2018puo}
\bibitem{Aaboud:2018puo} 
  M.~Aaboud {\it et al.} [ATLAS Collaboration],
  %``Constraints on off-shell Higgs boson production and the Higgs boson total width in $ZZ\to4\ell$ and $ZZ\to2\ell2\nu$ final states with the ATLAS detector,''
  Phys.\ Lett.\ B {\bf 786}, 223 (2018)
 % doi:10.1016/j.physletb.2018.09.048
  [arXiv:1808.01191 [hep-ex]].
  %%CITATION = doi:10.1016/j.physletb.2018.09.048;%%
  %8 citations counted in INSPIRE as of 10 Jan 2019
  
  %\cite{Sirunyan:2019twz}
\bibitem{Sirunyan:2019twz} 
  A.~M.~Sirunyan {\it et al.} [CMS Collaboration],
  %``Measurements of the Higgs boson width and anomalous HVV couplings from on-shell and off-shell production in the four-lepton final state,''
  arXiv:1901.00174 [hep-ex].
  %%CITATION = ARXIV:1901.00174;%%
  %12 citations counted in INSPIRE as of 25 May 2019
    
  %\cite{Falkowski:2013dza}
\bibitem{Falkowski:2013dza} 
  A.~Falkowski, F.~Riva and A.~Urbano,
  %``Higgs at last,''
  JHEP {\bf 1311}, 111 (2013)
%  doi:10.1007/JHEP11(2013)111
  [arXiv:1303.1812 [hep-ph]].
  %%CITATION = doi:10.1007/JHEP11(2013)111;%%
  %232 citations counted in INSPIRE as of 13 Feb 2019

\bibitem{ATLAS_mu}
  ATLAS Collaboration,
  ATL-PHYS-PUB-2018-054
  
 %\cite{CMS:2018qgz}
\bibitem{CMS:2018qgz} 
  CMS Collaboration,
  %``Sensitivity projections for Higgs boson properties measurements at the HL-LHC,''
  CMS-PAS-FTR-18-011.
  %%CITATION = CMS-PAS-FTR-18-011;%%
  %2 citations counted in INSPIRE as of 28 Feb 2019
  
\bibitem{ATLAS_width}
  ATLAS Collaboration,
  ATL-PHYS-PUB-2015-024
     
       %\cite{Perez:2015aoa}
\bibitem{Perez:2015aoa} 
  G.~Perez, Y.~Soreq, E.~Stamou and K.~Tobioka,
  %``Constraining the charm Yukawa and Higgs-quark coupling universality,''
  Phys.\ Rev.\ D {\bf 92}, no. 3, 033016 (2015)
%  doi:10.1103/PhysRevD.92.033016
  [arXiv:1503.00290 [hep-ph]].
  %%CITATION = doi:10.1103/PhysRevD.92.033016;%%
  %55 citations counted in INSPIRE as of 19 Feb 2019

%\cite{Perez:2015lra}
\bibitem{Perez:2015lra} 
  G.~Perez, Y.~Soreq, E.~Stamou and K.~Tobioka,
  %``Prospects for measuring the Higgs boson coupling to light quarks,''
  Phys.\ Rev.\ D {\bf 93}, no. 1, 013001 (2016)
  doi:10.1103/PhysRevD.93.013001
  [arXiv:1505.06689 [hep-ph]].
  %%CITATION = doi:10.1103/PhysRevD.93.013001;%%
  %48 citations counted in INSPIRE as of 07 May 2019

%\cite{Koenig:2015pha}
\bibitem{Koenig:2015pha} 
  M.~K\"{o}nig and M.~Neubert,
  %``Exclusive Radiative Higgs Decays as Probes of Light-Quark Yukawa Couplings,''
  JHEP {\bf 1508}, 012 (2015)
  doi:10.1007/JHEP08(2015)012
  [arXiv:1505.03870 [hep-ph]].
  %%CITATION = doi:10.1007/JHEP08(2015)012;%%
  %65 citations counted in INSPIRE as of 18 Apr 2019
  
  %\cite{Bodwin:2013gca}
\bibitem{Bodwin:2013gca} 
  G.~T.~Bodwin, F.~Petriello, S.~Stoynev and M.~Velasco,
  %``Higgs boson decays to quarkonia and the $H\bar{c}c$  coupling,''
  Phys.\ Rev.\ D {\bf 88}, no. 5, 053003 (2013)
  doi:10.1103/PhysRevD.88.053003
  [arXiv:1306.5770 [hep-ph]].
  %%CITATION = doi:10.1103/PhysRevD.88.053003;%%
  %110 citations counted in INSPIRE as of 18 Apr 2019  
   
 %\cite{Bodwin:2014bpa}
\bibitem{Bodwin:2014bpa} 
  G.~T.~Bodwin, H.~S.~Chung, J.~H.~Ee, J.~Lee and F.~Petriello,
  %``Relativistic corrections to Higgs boson decays to quarkonia,''
  Phys.\ Rev.\ D {\bf 90}, no. 11, 113010 (2014)
 % doi:10.1103/PhysRevD.90.113010
  [arXiv:1407.6695 [hep-ph]].
  %%CITATION = doi:10.1103/PhysRevD.90.113010;%%
  %44 citations counted in INSPIRE as of 07 Jan 2019
  
  
  
 %\cite{Aaboud:2018txb}
\bibitem{Aaboud:2018txb} 
  M.~Aaboud {\it et al.} [ATLAS Collaboration],
  %``Searches for exclusive Higgs and $Z$ boson decays into $J/\psi\gamma$, $\psi(2S)\gamma$, and $\Upsilon(nS)\gamma$ at $\sqrt{s}=13$ TeV with the ATLAS detector,''
  Phys.\ Lett.\ B {\bf 786}, 134 (2018)
%  doi:10.1016/j.physletb.2018.09.024
  [arXiv:1807.00802 [hep-ex]].
  %%CITATION = doi:10.1016/j.physletb.2018.09.024;%%
  %4 citations counted in INSPIRE as of 09 Jan 2019
  
    %\cite{Alwall:2014hca}
\bibitem{Alwall:2014hca} 
  J.~Alwall {\it et al.},
  %``The automated computation of tree-level and next-to-leading order differential cross sections, and their matching to parton shower simulations,''
  JHEP {\bf 1407}, 079 (2014)
%  doi:10.1007/JHEP07(2014)079
  [arXiv:1405.0301 [hep-ph]].
  %%CITATION = doi:10.1007/JHEP07(2014)079;%%
  %3494 citations counted in INSPIRE as of 19 Mar 2019
  
\bibitem{Aaboud:}
  M.~Aaboud {\it et al.} [ATLAS Collaboration],
  %``Search for the Standard Model Higgs and Z Boson decays to J/? ?: HL-LHC projections, 
  Tech.\ Rep.\ ATL-PHYS-PUB-2015-043, CERN, Geneva, Sep, 2015.
  
%\cite{Bodwin:2016edd}
\bibitem{Bodwin:2016edd} 
  G.~T.~Bodwin, H.~S.~Chung, J.~H.~Ee and J.~Lee,
  %``New approach to the resummation of logarithms in Higgs-boson decays to a vector quarkonium plus a photon,''
  Phys.\ Rev.\ D {\bf 95}, no. 5, 054018 (2017)
  doi:10.1103/PhysRevD.95.054018
  [arXiv:1603.06793 [hep-ph]].
  %%CITATION = doi:10.1103/PhysRevD.95.054018;%%
  %10 citations counted in INSPIRE as of 17 Apr 2019  
  
  %\cite{Brivio:2015fxa}
\bibitem{Brivio:2015fxa} 
  I.~Brivio, F.~Goertz and G.~Isidori,
  %``Probing the Charm Quark Yukawa Coupling in Higgs+Charm Production,''
  Phys.\ Rev.\ Lett.\  {\bf 115}, no. 21, 211801 (2015)
 % doi:10.1103/PhysRevLett.115.211801
  [arXiv:1507.02916 [hep-ph]].
  %%CITATION = doi:10.1103/PhysRevLett.115.211801;%%
  %28 citations counted in INSPIRE as of 12 Feb 2019
  
    %\cite{Sirunyan:2017ezt}
%\bibitem{Sirunyan:2017ezt} 
%  A.~M.~Sirunyan {\it et al.} [CMS Collaboration],
  %``Identification of heavy-flavour jets with the CMS detector in pp collisions at 13 TeV,''
%  JINST {\bf 13}, no. 05, P05011 (2018)
 % doi:10.1088/1748-0221/13/05/P05011
%  [arXiv:1712.07158 [physics.ins-det]].
  %%CITATION = doi:10.1088/1748-0221/13/05/P05011;%%
  %243 citations counted in INSPIRE as of 03 Apr 2019
  

  
   %\cite{Aaboud:2018fhh}
\bibitem{Aaboud:2018fhh} 
  M.~Aaboud {\it et al.} [ATLAS Collaboration],
  %``Search for the Decay of the Higgs Boson to Charm Quarks with the ATLAS Experiment,''
  Phys.\ Rev.\ Lett.\  {\bf 120}, no. 21, 211802 (2018)
%  doi:10.1103/PhysRevLett.120.211802
  [arXiv:1802.04329 [hep-ex]].
  %%CITATION = doi:10.1103/PhysRevLett.120.211802;%%
  %18 citations counted in INSPIRE as of 17 Jan 2019

\bibitem{ATLAS:ZhUpgrade}  
  ATLAS Collaboration, 
  ATL-PHYS-PUB-2018-016

\bibitem{CMStwiki}
   CMS Collaboration,
   CMS-DP-2018-046.
       
  %\cite{Yu:2016rvv}
\bibitem{Yu:2016rvv} 
  F.~Yu,
  %``Phenomenology of Enhanced Light Quark Yukawa Couplings and the $W^\pm h$ Charge Asymmetry,''
  JHEP {\bf 1702}, 083 (2017)
%  doi:10.1007/JHEP02(2017)083
  [arXiv:1609.06592 [hep-ph]].
  %%CITATION = doi:10.1007/JHEP02(2017)083;%%
  %18 citations counted in INSPIRE as of 15 Mar 2019
  
  %\cite{Bishara:2016jga}
\bibitem{Bishara:2016jga} 
  F.~Bishara, U.~Haisch, P.~F.~Monni and E.~Re,
  %``Constraining Light-Quark Yukawa Couplings from Higgs Distributions,''
  Phys.\ Rev.\ Lett.\  {\bf 118}, no. 12, 121801 (2017)
  doi:10.1103/PhysRevLett.118.121801
  [arXiv:1606.09253 [hep-ph]].
  %%CITATION = doi:10.1103/PhysRevLett.118.121801;%%
  %53 citations counted in INSPIRE as of 18 Jun 2019
  
  %\cite{Soreq:2016rae}
\bibitem{Soreq:2016rae} 
  Y.~Soreq, H.~X.~Zhu and J.~Zupan,
  %``Light quark Yukawa couplings from Higgs kinematics,''
  JHEP {\bf 1612}, 045 (2016)
  doi:10.1007/JHEP12(2016)045
  [arXiv:1606.09621 [hep-ph]].
  %%CITATION = doi:10.1007/JHEP12(2016)045;%%
  %35 citations counted in INSPIRE as of 18 Jun 2019
  
%\cite{Bonner:2016sdg}
\bibitem{Bonner:2016sdg} 
  G.~Bonner and H.~E.~Logan,
  %``Constraining the Higgs couplings to up and down quarks using production kinematics at the CERN Large Hadron Collider,''
  arXiv:1608.04376 [hep-ph].
  %%CITATION = ARXIV:1608.04376;%%
  %9 citations counted in INSPIRE as of 18 Jun 2019
  
    %\cite{Sirunyan:2018sgc}
\bibitem{Sirunyan:2018sgc} 
  A.~M.~Sirunyan {\it et al.} [CMS Collaboration],
  %``Measurement and interpretation of differential cross sections for Higgs boson production at $\sqrt{s} =$ 13 TeV,''
  Phys.\ Lett.\ B {\bf 792}, 369 (2019)
  doi:10.1016/j.physletb.2019.03.059
  [arXiv:1812.06504 [hep-ex]].
  %%CITATION = doi:10.1016/j.physletb.2019.03.059;%%
  %6 citations counted in INSPIRE as of 18 Jun 2019


 
  

  
\end{thebibliography}
\end{document}